\documentclass[a4paper,twoside,openany,11pt]{memoir} 
\pdfoutput=1

\def\fullheadfoot{0} 
\usepackage[english]{babel}
\usepackage{microtype,xspace}
\usepackage[utf8]{inputenc}	

\usepackage{amsfonts,amsmath,amssymb,bm,mathrsfs,mathtools,dsfont} 	
\allowdisplaybreaks 	
\usepackage[table]{xcolor}
\usepackage{hyperref}
\usepackage{slashed}
\usepackage{multicol}

\usepackage{siunitx}
\usepackage{geometry}
\usepackage[sort&compress,numbers,merge]{natbib}
\usepackage{chapterbib}
\addto\captionsenglish{\renewcommand*{\bibname}{References}}
\makeatletter
\renewcommand{\@memb@bchap}{ 
\bibmark \prebibhook
}
\makeatother

\usepackage{graphicx}
\usepackage{tikz}
\graphicspath{{./Figures/}}
\usepackage{caption,subcaption}
\usepackage{multirow}
\renewcommand{\arraystretch}{1.2}
\usepackage{tabularx}
\newcolumntype{Y}{>{\centering\arraybackslash}X}

\usepackage[shortlabels]{enumitem}
\setlist{itemsep=.1em,topsep=.5em}
\SetEnumerateShortLabel{i}{\textit{\roman*}}

\definecolor{red}{rgb}{0.6,.0706,.1373}
\definecolor{blue}{rgb}{0,0.396,0.741}
\definecolor{green}{rgb}{0.25,0.6,0.2}
\definecolor{teal}{rgb}{0.11,0.6,0.6}
\definecolor{orange}{rgb}{.8, .4806, 0.173}
\definecolor{yellow}{rgb}{.8, .7, 0.05}
\colorlet{blueref}{blue!80!black}
\colorlet{bluelink}{blue!90!black}
\hypersetup{
	colorlinks, 
	bookmarksopen, 
	bookmarksnumbered,
	citecolor=blueref, 		
	linkcolor=bluelink,	
	urlcolor=bluelink,			
	pdftitle={Two-loop running in the bosonic SMEFT using functional methods},    
	pdfauthor={Born, Fuentes-Martín, Kvedaraitė, and Thomsen},    	
}

\addto\captionsenglish{}

\renewcommand{\contentsname}{Contents}
\renewcommand{\printtoctitle}[1]{}

\settocdepth{subsection}
\makeatletter
\newcommand*\ifthispageodd{%
  \checkoddpage
  \ifoddpage
    \expandafter\@firstoftwo
  \else
    \expandafter\@secondoftwo
  \fi
}
\makeatother

\usepackage[noindentafter]{titlesec} 
\counterwithout{section}{chapter} 
\counterwithout{figure}{chapter} 
\counterwithout{table}{chapter} 
\numberwithin{equation}{section} 
\setsecnumdepth{subsubsection}

\DeclareMathVersion{sans}
\SetSymbolFont{operators}{sans}{OT1}{cmbr}{m}{n}
\SetSymbolFont{letters}{sans}{OML}{cmbrm}{m}{it}
\SetSymbolFont{symbols}{sans}{OMS}{cmbrs}{m}{n}
\SetMathAlphabet{\mathit}{sans}{OT1}{cmbr}{m}{sl}
\SetMathAlphabet{\mathbf}{sans}{OT1}{cmbr}{bx}{n}
\SetMathAlphabet{\mathtt}{sans}{OT1}{cmtl}{m}{n}
\SetSymbolFont{largesymbols}{sans}{OMX}{iwona}{m}{n}

\DeclareMathVersion{boldsans}
\SetSymbolFont{operators}{boldsans}{OT1}{cmbr}{b}{n}
\SetSymbolFont{letters}{boldsans}{OML}{cmbrm}{b}{it}
\SetSymbolFont{symbols}{boldsans}{OMS}{cmbrs}{b}{n}
\SetMathAlphabet{\mathit}{boldsans}{OT1}{cmbr}{b}{sl}
\SetMathAlphabet{\mathbf}{boldsans}{OT1}{cmbr}{bx}{n}
\SetMathAlphabet{\mathtt}{boldsans}{OT1}{cmtl}{b}{n}
\SetSymbolFont{largesymbols}{boldsans}{OMX}{iwona}{bx}{n}

\titleformat{\section}{\centering \needspace{5\baselineskip}\Large \bfseries \sffamily \mathversion{boldsans} \color{blue!80!black} }{\thesection}{15pt}{}{}
\titlespacing{\section}{0pt}{15pt}{5pt}
\titleformat{\subsection}{\large \sffamily \mathversion{sans} \color{blue!70!black} }{\thesubsection}{10pt}{}{}
\titlespacing{\subsection}{0pt}{10pt}{5pt}
\titleformat{\subsubsection}{\normalsize \sffamily \itshape \mathversion{sans} \color{blue!70!black} }{\thesubsubsection}{10pt}{}{}
\titlespacing{\subsubsection}{0pt}{10pt}{0pt}

\newcommand{\sectionlike}[1]{\phantomsection \addcontentsline{toc}{section}{#1} \setcounter{subsection}{0} \sectionmark{#1}
		\begin{center}
		\needspace{5\baselineskip}
		\Large \bfseries \sffamily \mathversion{boldsans} \color{blue!80!black} #1  
		\end{center}
	\vspace{-5pt} 
}

\ifnum\fullheadfoot=1
	\makepagestyle{standardstyle}
	\makeoddhead{standardstyle}{\sffamily \mathversion{subsectionmath} \rightmark}{}{\sffamily \bfseries{\thepage}}
	\makeevenhead{standardstyle}{\sffamily \bfseries{\thepage}}{}{\sffamily \mathversion{subsectionmath} \leftmark}
	\makeheadrule{standardstyle}{\textwidth}{\normalrulethickness}
	\makepsmarks{standardstyle}{
		\nouppercaseheads
		\createmark{section}{both}{shownumber}{\ }{\hspace{1.2em}  }
		\createmark{subsection}{right}{shownumber}{}{\hspace{1.2em} }
	}
	
	\copypagestyle{appstyle}{standardstyle}
	\makeevenhead{appstyle}{\sffamily \bfseries{\thepage}}{}{ \sffamily \mathversion{subsectionmath} \leftmark}
	\makepsmarks{appstyle}{
		\nouppercaseheads
		\createmark{section}{both}{nonumber}{\ }{\ \ }
		\createmark{subsection}{right}{shownumber}{}{\ \ }
	}
\else 
	\makepagestyle{standardstyle}
	\makeoddfoot{standardstyle}{}{\sffamily -- {\bfseries\thepage} --}{} 
	\makeevenfoot{standardstyle}{}{\sffamily -- {\bfseries\thepage} --}{}
	\copypagestyle{appstyle}{standardstyle}
\fi

\createplainmark{toc}{both}{\contentsname}
\createplainmark{bib}{both}{\bibname}

\makeatletter
\let\MyIntOrig\int
\def\MyIntSpace{\hspace{-.35em}} 
\def\int{\MyInt}
\def\MyInt{\MyIntOrig\MyIntSkipMaybe}
\def\MyIntSkipMaybe{
	\@ifnextchar_{\MyIntSkipScript}{%
		\@ifnextchar^{\MyIntSkipScript}{%
			\@ifnextchar\limits{\MyIntSkipTok}{%
				\@ifnextchar\nolimits{\MyIntSkipTok}{%
					\MyIntSpace}}}}%
}
\def\MyIntSkipScript#1#2{#1{#2}\MyIntSkipMaybe}
\def\MyIntSkipTok#1{#1\MyIntSkipMaybe}

\newcommand{\pushright}[1]{\ifmeasuring@#1\else\omit\hfill$\displaystyle#1$\fi\ignorespaces}
\makeatother

\newcommand{\Tr}{\mathop{\mathrm{Tr}}}

\newcommand{\eminus}{\vcenter{\hbox{\scalebox{0.6}[1]{$ - $}}}}	
\newcommand{\ord}[1]{\mathcal{O}( #1 )}

\newcommand{\dd}{\mathop{}\!\mathrm{d}}
\newcommand{\ud}[2]{\phantom{}^{#1}\phantom{}_{#2}}

\newcommand{\ineqgraphics}[1]{\vcenter{\hbox{\includegraphics[]{#1}}}}
\newcommand{\sscript}[1]{{\scriptscriptstyle \mathrm{#1}}}

\renewcommand{\L}{\mathcal{L}}

\newcommand{\SU}{\mathrm{SU}}

\newcommand{\ctop}{\boldsymbol{\Delta}}
\newcommand{\kop}{\boldsymbol{K}}
\newcommand{\rstaroperation}{$ \boldsymbol{R}^{\ast} $-operation\xspace}

\newcommand{\cO}{\mathcal{O}}

\newcommand{\lzm}{\left(}
\newcommand{\dzm}{\right)}

\newcommand{\bef}{$ \beta $-function\xspace}
\newcommand{\befs}{$ \beta $-functions\xspace}
\newcommand{\msbar}{$ \overline{\text{\small MS}} $\xspace}

\colorlet{redref}{red!80!violet}

\begin{document}

\thispagestyle{empty}
\renewcommand*{\thefootnote}{\fnsymbol{footnote}}

\begin{center}
    {\sffamily \bfseries \fontsize{16.}{20}\selectfont \mathversion{boldsans}
    Two-Loop Running in the Bosonic SMEFT\\ Using Functional Methods\\[-.5em]
    \textcolor{blue!80!black}{\rule{.45\textwidth}{2pt}}\\
    \vspace{.03\textheight}}
    {\sffamily \mathversion{sans} \Large 
    Lukas Born,$^{1}$\footnote{lukas.born@unibe.ch}
    Javier Fuentes-Martín,$^{2}$\footnote{javier.fuentes@ugr.es} 
    Sandra Kvedaraitė,$^{2}$\footnote{skvedaraite@ugr.es}\\ 
    and Anders Eller Thomsen$^{1}$\footnote{anders.thomsen@unibe.ch}
    }\\[1.25em]
    { \small \sffamily \mathversion{sans} 
        $^{1}\,$Albert Einstein Center for Fundamental Physics, Institute for Theoretical Physics, University of Bern, CH-3012 Bern, Switzerland\\[5pt] 
        $^{2}\,$Departamento de Física Teórica y del Cosmos, Universidad de Granada,\\
        Campus de Fuentenueva, E–18071 Granada, Spain
    }
    \\[.005\textheight]{\itshape \sffamily \today}
    \\[.03\textheight]
\end{center}
\setcounter{footnote}{0}
\renewcommand*{\thefootnote}{\arabic{footnote}}%
\suppressfloats	

\begin{abstract}\vspace{+.01\textheight}
    The next goalpost in precision calculations for physics beyond the Standard Model is determining the two-loop renormalization group (RG) equations in the Standard Model Effective Field Theory (SMEFT). We progress towards this goal by determining the RG equations for a simplified version of the SMEFT without any fermion fields. Our calculation relies on functional methods that are newly developed for multi-loop computations and adapted for the determination of RG equations here. 
\end{abstract}

\newpage


\section{Introduction}
It seems increasingly likely that new physics (NP)---whatever it might be---is too heavy to be produced on-shell at the LHC or, alternatively, too feebly interacting to have been observed as of yet. Pursuing the former hypothesis, one of the most promising strategies to search for NP consists in looking for its indirect traces in precision observables, following the great tradition of indirect particle discoveries stretching back to the inferences of the existence of neutrinos, the charm quark, and the third Standard Model (SM) family. This program puts strong demands on theoretical SM predictions but also, increasingly, on the precision of beyond-the-SM (BSM) calculations. Indeed, there are many effects that can only be captured once higher-order corrections are considered. For instance, electromagnetic dipoles are exclusively generated at the loop level; RG mixing of operators have been found to give non-trivial contributions to observables that were at first believed to be unrelated~\cite{Feruglio:2016gvd,Feruglio:2017rjo,Crivellin:2018yvo}; and one-loop matching corrections have been shown to play a crucial role in the phenomenology of many NP models, see e.g.~\cite{Gherardi:2020qhc,Crivellin:2022fdf,Guedes:2022cfy,Crivellin:2023xbu,Lizana:2023kei,DasBakshi:2024krs,Cepedello:2024ogz}.

The precision program, however, does not end at one-loop order.\footnote{In a related effort, there is progress towards determining the one-loop running of the SMEFT at dimension eight~\cite{Chala:2021pll,DasBakshi:2022mwk,DasBakshi:2023htx,Chala:2023xjy,Bakshi:2024wzz,Liao:2024xel}.}
CP-odd triple-gauge interactions are generated only at two-loop order~\cite{Bakshi:2021ofj}, while the top Yukawa coupling is large enough to produce important running effects beyond one-loop order~\cite{Ardu:2021koz,Allwicher:2023aql}. Furthermore, the scheme independence of predictions involving one-loop matching calculations can be obtained only with the inclusion of two-loop running~\cite{Ciuchini:1993ks,Ciuchini:1993fk}. The effects of heavy NP are commonly captured in the SM Effective Field Theory (SMEFT)~\cite{Buchmuller:1985jz,Grzadkowski:2010es}, which has by now become the predominant framework for BSM searches given its model comprehensiveness. The one-loop RG equations for the dimension-six SMEFT have been known for more than a decade~\cite{Jenkins:2013zja,Jenkins:2013wua,Alonso:2013hga}. However, in light of this discussion, it is timely to work toward the determination of the RG equations at two-loop order (a few partial results have already appeared in~\cite{Jenkins:2023bls,DiNoi:2024ajj}). In this paper, we take a first leap in this direction by computing the two-loop RG equations for the bosonic version of the dimension-six SMEFT.

The full calculation of two-loop running in the SMEFT, even when restricted to the bosonic sector, is a formidable task already at dimension six. Firstly, the sheer number of operators poses a serious organizational challenge compared to typical renormalizable models. Secondly, the evaluation of diagrams is significantly more involved than at one-loop order, requiring careful management of subdivergences and IR rearrangement. The first of these challenges can be ameliorated with the use of functional methods. Indeed, the renaissance of these methods for RG and matching calculations in the last decade has demonstrated their utility in organizing such calculations in an efficient manner~\cite{Henning:2014wua,Ellis:2016enq,Fuentes-Martin:2016uol,Zhang:2016pja,Henning:2016lyp,Kramer:2019fwz,Cohen:2020fcu,Fuentes-Martin:2020udw,Cohen:2020qvb,Dittmaier:2021fls,Fuentes-Martin:2022jrf}. In particular, the use of functional methods lets us maintain manifest gauge covariance through all intermediate steps, and the loop calculation gives the counterterm Lagrangian directly as an output rather than merely giving the divergent part of an amplitude. Recently, these methods have been extended to the two-loop order for pure scalar theories~\cite{Fuentes-Martin:2023ljp}, with their generalization to arbitrary theories presented in~\cite{Fuentes-Martin:2024agf}. In this paper, we solve the second challenge by adapting a local version of the $ \boldsymbol{R}^{\ast} $-operation~\cite{Herzog:2017bjx} to subtract all subdivergences in the two-loop Feynman integrals. This operation serves as an ideal complement to the functional tools when extracting UV divergences. For our calculations, we will make use of a bespoke computer pipeline---built on the \texttt{Matchete} package~\cite{Fuentes-Martin:2022jrf}---that fully automates the determination of two-loop \msbar counterterms in bosonic theories. It is our intent to extend these methods to fermions in a follow-up work, facilitating the calculations of the RG functions of the full SMEFT at dimension six,\footnote{The inclusion of fermions involves additional challenges related to the non-trivial continuation of the Dirac algebra to $ d $ dimensions, which invalidates the usual four-dimensional identities and calls for careful handling of evanescent operators~\cite{Dugan:1990df,Buras:1989xd,Herrlich:1994kh,Fuentes-Martin:2022vvu}, influencing the flow of the physical operators.} and, eventually, incorporate these computational developments into a public version of the \texttt{Matchete} package. 

This paper is organized as follows. In Section~\ref{sec:methods}, we briefly outline the two-loop functional methods. We further show how to implement the $ \boldsymbol{R}^{\ast} $ operation to extract the counterterm Lagrangian directly and how to account for symmetries to simplify the calculation. In Section~\ref{sec:results}, we present the bosonic SMEFT and our results. We discuss the details of our computer implementation and how we have validated the code. Section~\ref{sec:conclusion} concludes with a brief outlook. The two-loop \befs of the bosonic SMEFT can be found in Appendix~\ref{app:betas} and in the ancillary file \texttt{betaFunctions.m}.

\section{Methods} \label{sec:methods}
We begin by summarizing the essentials of the functional methods at two-loop order  
before proceeding with their practical application to counterterm calculations.

\subsection{Functional evaluation of the quantum effective action}
\begin{figure*}[t]
    \includegraphics[width=.9\textwidth]{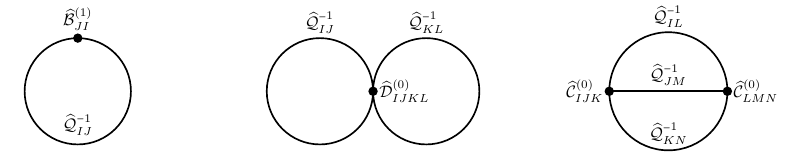}
     \caption{Graphical representation of the terms appearing in the two-loop effective action.}
     \label{fig:eff_action}
\end{figure*}

Functional methods are techniques for calculating the QFT path integral through a saddlepoint (loop) approximation all at once rather than focusing on one specific Green's function at a time~\cite{Coleman:1973jx,Jackiw:1974cv,Iliopoulos:1974ur,Bijnens:1999hw,Fuentes-Martin:2023ljp}. In this expansion, one ends up with but a few generic supergraphs (topologies), such as those shown in Fig.~\ref{fig:eff_action}, dressed with arbitrary background field insertions. It is then possible to evaluate the graphs in a manifestly gauge-invariant manner accounting for summation over all fields in the theory, effectively calculating the whole generating functional all at once.  

To this end, one expands the action around a stationary point:
given a generic action $ S[\eta] $ of fields collectively denoted by $ \eta_I\equiv \eta_a(x) $, we parameterize the expansion around the background field configurations $ \hat{\eta}_I $ by 
    \begin{equation} \label{eq:action_expansion}
    S[\eta + \hat{\eta}] = \widehat{S}
    + \sum_{\ell=0}^{\infty} \hbar^\ell \Big[ \eta_I \widehat{\mathcal{A}}^{(\ell)}_I + \tfrac{1}{2} \eta_I \eta_J \widehat{\mathcal{B}}^{(\ell)}_{IJ} 
    +\tfrac{1}{6} \eta_I \eta_J \eta_K \widehat{\mathcal{C}}^{(\ell)}_{IJK} +\tfrac{1}{24} \eta_I \eta_J \eta_K \eta_L \widehat{\mathcal{D}}^{(\ell)}_{IJKL} +\ord{\eta^5} \Big], 
    \end{equation}
where, e.g., $ \widehat{\mathcal{B}}^{(\ell)}_{IJ} $ abbreviates the functional dependence $ \mathcal{B}^{(\ell)}_{IJ}[\hat{\eta}] $, and the superscript $ (\ell) $ refers to the loop order of the quantity. Throughout this section, we use DeWitt notation with implicit coordinate integration for the repeated indices; for instance, $ \eta_I \eta_J \widehat{\mathcal{B}}^{(\ell)}_{IJ}  \equiv \int_{x,y} \, \eta_a(x) \eta_b(y) \widehat{\mathcal{B}}^{(\ell)}_{ab}(x,y) $.\footnote{We use shorthand notation for spacetime integrals: $\int_x \equiv \int \dd^d x$.}
It is conventional to refer to the kinetic (fluctuation) operator as $ \mathcal{Q}_{IJ} \equiv \mathcal{B}^{(0)}_{IJ} $, since it plays a special role as the propagator. 

The functional formalism provides methods for determining the quantum effective action, the generating functional for one-particle-irreducible (1PI) Green's functions, by performing a Gaussian approximation of the path integral around the background fields. We refer the reader to~\cite{Fuentes-Martin:2024agf} for details on the fully general case and merely list the main results here. 
At two-loop order the effective action reads\footnote{The ghost fields resulting from gauge-fixing are Grassmann odd and their inclusion introduces an overall sign in some contractions, which are explicitly shown in~\cite{Fuentes-Martin:2024agf}. These Grassmann signs are analogous to those in regular Feynman graphs but for the functional diagrams in Figure~\ref{fig:eff_action}.}
    \begin{align} \label{eq:effective_action}
    \Gamma[\hat{\eta}] &= \widehat{S} + \dfrac{i\hbar}{2} \Tr \log \widehat{\mathcal{Q}} + \dfrac{i\hbar^2}{2} \widehat{\mathcal{Q}}_{IJ}^{\eminus1} \widehat{\mathcal{B}}_{JI}^{(1)}
    -  \dfrac{\hbar^2}{8} \widehat{\mathcal{D}}^{(0)}_{IJKL} \widehat{\mathcal{Q}}_{IJ}^{\eminus1} \widehat{\mathcal{Q}}_{KL}^{\eminus1}
    + \dfrac{\hbar^2}{12} \widehat{\mathcal{C}}_{IJK}^{(0)} \widehat{\mathcal{Q}}_{IL}^{\eminus1} \widehat{\mathcal{Q}}_{JM}^{\eminus1} \widehat{\mathcal{Q}}_{KN}^{\eminus1} \widehat{\mathcal{C}}^{(0)}_{LMN} \nonumber \\* 
    &= \widehat{S} + \dfrac{i\hbar}{2}  G_\mathrm{log} + \dfrac{i\hbar^2}{2} G_\mathrm{ct.} - \dfrac{\hbar^2}{8} G_\mathrm{f8.} + \dfrac{\hbar^2}{12} G_\mathrm{ss.},
    \end{align}
where we use explicit powers of $\hbar$ as a loop-counting parameter. The action $\widehat{S}$ contains the tree-level classical action as well as the counterterms of the theory, which in the MS (or \msbar) scheme are divergent terms fixed to cancel the UV-divergent piece of the other contributions to the effective action. Apart from the counterterms, the one-loop contributions to the effective action appear functionally in the form of a log term, whereas the two-loop contributions consist of three terms, illustrated in Fig.~\ref{fig:eff_action}: a one-loop counterterm insertion in a one-loop topology, the figure-8 topology, and the sunset topology. All terms are cast as tensorial DeWitt contractions of the generic indices, which labels both internal degrees of freedom and spacetime points.
The effective action is generically non-local and cannot be evaluated directly: it is the sum of all one- and two-loop 1PI Green's functions, which cannot be expressed in a closed form. However, the UV divergent parts of these terms are local and can be determined from a convenient expansion.

The first step towards obtaining an expansion of the effective action is to cast the expression as familiar Feynman integrals in momentum space. These techniques have been developed for both single~\cite{Aitchison:1984ys,Fraser:1984zb,Henning:2014wua} (among many others) and multi-loop computations~\cite{Fuentes-Martin:2023ljp,Fuentes-Martin:2024agf}. An advantage of the functional approach as we proceed to gauge theories is that one can maintain explicit (background) gauge invariance at all stages of the calculation. 
As shown in~\cite{Fuentes-Martin:2024agf}, this is possible at multi-loop order with the introduction of a covariant version of the delta function:
\begin{equation} \label{eq:cov_delta_definition}
    \dfrac{\delta \eta_a(x)}{ \delta \eta_b(y) } = \delta_{ab}(x,\, y), \qquad \delta_{ab}(x,\, y) \equiv U_{ab} (x,\, y) \delta(x-y),
\end{equation}
where $U(x,y)$ is a Wilson line responsible for the parallel transport of the gauge configurations from the point $y$ to $x$, telling us how to compare fields at different spacetime points. For simplicity, we assume that the Wilson line follows a straight line from $x$ to $y$, as this provides several useful properties that facilitate practical calculations~\cite{Barvinsky:1985an,Kuzenko:2003eb}. In effect, the only property that will be applied in our calculations is the expansion of the covariant derivative acting on a Wilson line:
\begin{equation} \label{eq:D_on_U_right}
    D^\nu_x U(x,y) = i \sum_{n=1}^{\infty} \dfrac{(\eminus 1)^n}{(n+1)!} (x-y)_{\underline{n}} \big[D_x^{\underline{n-1}} G^{\mu_n \nu}(x) \big] U(x,y),
\end{equation} 
along with $ U(x,y) U(y,x)= U(x, x) = \mathds{1} $. We employ indices with underscores as short-hand notation for contractions of multiple Lorentz indices, e.g., 
    \begin{equation}
    (x-y)_{\underline{n}} D_x^{\underline{n-1}} = (x-y)_{\mu_1} \cdots (x-y)_{\mu_n} D_x^{\mu_1} \cdots D_x^{\mu_{n-1}}.
    \end{equation}

Returning to the action~\eqref{eq:action_expansion}, its locality ensures that the two-, three-, and four-point functions are differential operators acting on covariant delta functions: 
    \begin{align}
    \widehat{\mathcal{Q}}_{IJ} & = Q_{ac}(x,\,P_x) \delta\ud{c}{b}(x,y), \\ 
    \widehat{\mathcal{C}}^{(0)}_{IJK} &= \sum_{m,n=0} \Big[ C^{(\underline{m},\underline{n})}_{abc}(x) P_y^{\underline{m}} P_z^{\underline{n}} \Big] \delta\ud{d}{b}(x,y)\delta\ud{e}{c}(x,z), \\
    \widehat{\mathcal{D}}^{(0)}_{IJKL} &= \sum_{m,n,r=0} \! \Big[D^{(\underline{m},\underline{n},\underline{r})}_{abcd}(x) P_y^{\underline{m}} P_z^{\underline{n}} P_w^{\underline{r}} \Big]  \delta\ud{e}{b}(x,y) \delta\ud{f}{c}(x,z) \delta\ud{g}{d}(x,w) ,
    \end{align}
where the covariant momentum operators are given by $ P_x^\mu = i D_x^\mu $ in terms of the ordinary covariant derivatives in position space. 
This parametrization lets us cast the sunset and figure-8 contributions to the two-loop effective action as momentum-space Feynman integrals~\cite{Fuentes-Martin:2024agf}:
    \begin{align}\label{eq:Gss}
    G_\mathrm{ss.} = \!\! \sum_{m,n,m',n'} \!\!(\eminus 1)^{m+n} & \sum_{r,s,t,u=0}^{\infty} \dfrac{i^r (\eminus i)^{s+t+u}}{r!} \int_{x} \, \int_{k\ell q } \delta(q+k+\ell) C^{(\underline{m}, \underline{n})}_{abc}(x) D_x^{\underline{r}} C^{(\underline{m}', \underline{n}')}_{a'b'c'}(x) \nonumber\\
    &\times \big[\partial_q^{{\underline{r}+ \underline{s}+ \underline{t} + \underline{u}}} Z^{(0,0,\underline{s})}_{aa'}(x,q) \big]\,
    Z^{(\underleftarrow{m}, \underline{m}', \underline{t})}_{bb'}(x,k) \,
    Z^{(\underleftarrow{n}, \underline{n}', \underline{u})}_{cc'}(x,\ell),
    \end{align}
and 
    \begin{equation}\label{eq:Gf8}
    G_\mathrm{f8.} =\sum_{m,n,r=0} (\eminus 1)^{n}\! \int_{x}\, \int_{k\ell}  D^{(\underline{m}, \underline{n}, \underline{r})}_{abcd}(x) Z_{ab}^{(0,\underline{m},0)}(x,k) Z_{cd}^{(\underleftarrow{n},\underline{r},0)}(x,\ell),
    \end{equation}
with momentum integrals $\int_k\, \equiv (2\pi)^{\eminus d}\! \int\, \dd^d k$. The left arrow under some of the $ m, n $ indices indicates that the order of these indices is reversed. 
This formulation might still look somewhat cryptic, as the inverse kinetic operator dressed with derivatives from the vertex has been parametrized as a Taylor series
    \begin{equation} \label{eq:def_Z}
    \sum^{\infty}_{s=0} (x-x')_{\underline{s}} Z^{(\underline{m}, \underline{m}', \underline{s}) }_{aa'}(x, k) \equiv \big[(P_x+k)^{\underline{m}} Q_{ab}^{\eminus 1}(x,\,P_x +k) (P_x+k)^{\underline{m}'} U\ud{b}{b'}(x,x') \big] U\ud{b'}{a'}(x',x). 
    \end{equation} 
This form has the benefit that explicit gauge invariance is manifest through the presence of the Wilson lines stemming from the covariant Dirac delta functions. 

We will see in the next section that the divergence of the effective action may be obtained from a Taylor expansion of the integrand (similar to the hard-region expansion used in matching calculations). Thus, we get a handle on the inverse kinetic operators by expanding in large loop momenta with all external fields and covariant derivatives taken to be small. After the expansion, 
$ Z^{(\underline{m}, \underline{m}', \underline{s}) }_{aa'}(x, k) $ is evaluated by commuting all derivatives through to the right to act on $U\ud{b}{b'}(x,x') $ in Eq.~\eqref{eq:def_Z} and then repeatedly applying Eq.~\eqref{eq:D_on_U_right} where in the end $ U(x,y) $ cancels against $ U(y,x)$. It is straightforward to determine where to truncate the expansion. Renormalizing a dimension-six EFT, we can focus on operators of mass-dimension no greater than six. The canonical dimension of the counterterms comes from background fields, covariant derivatives, and, potentially, coupling parameters. Thus, we simply need to truncate the expansion at mass-dimension six, not counting the loop momenta nor the dimension of the Wilson coefficients.

\subsection{The \texorpdfstring{$ R^{\ast} $}{R*}-method}\label{sec:RStar}

The UV divergences of one-loop Feynman integrals can be extracted rather easily by Taylor expanding the integrals around zero external momenta and masses. The UV divergence of the resulting integral is cleanly separated from the IR divergence and easily found by, e.g., the insertion of auxiliary masses by way of IR rearrangement.  
This useful trick is, unfortunately, no longer directly applicable at higher-loop order, as a more complex UV and IR structure emerges: 
Both UV and IR subdivergences give non-analytic contributions to the divergence and must be carefully subtracted. Additionally, Taylor expansion of the integrals produces new spurious IR divergences and therefore does not commute with the pole-operator. Several methods to overcome these challenges are used in the literature, but a local version of the $\boldsymbol{R^\ast}$-method is ideal for our purposes. The overview given here is somewhat sparse and we refer the reader to Ref.~\cite{Herzog:2017bjx} for further details. 

The $ \boldsymbol{R}^{\ast} $-method is an algorithm to disentangle 
and subtract all UV and IR divergences of a Feynman diagram $G$. More precisely, the $ \boldsymbol{R}^{\ast} $-operation renders $ G $ finite. It (or more conveniently the $\Bar{\boldsymbol{R}}^{\ast}$-operation) is defined on the graph $G$ by
    \begin{align}\label{eq:BarRStarOperation}
    \boldsymbol{R}^{\ast} G  = \Delta_\sscript{UV}(G) + \Bar{\boldsymbol{R}}^{\ast} G, \qquad \Bar{\boldsymbol{R}}^{\ast} =  \sum_{\substack{S \in \overline{W}_\sscript{UV}(G) \\ S' \in W_\sscript{IR}(G) \\ S \cap S' = \emptyset}} \Delta_\sscript{IR}(S') \ast \Delta_\sscript{UV}(S) \ast G/S \setminus S' \, .
    \end{align}
The sum runs over all UV-divergent proper subgraphs (including the empty set) $S$~\footnote{These are either UV divergent 1PI subgraphs without external legs or multiple disjoint such subgraphs.} and
IR subgraphs $S'$ (also including the empty set), the details of which are irrelevant to our application. Schematically, the remaining graph $G/S$ is constructed by contracting each disjoint component of $S$ in $G$ to a point.
The UV counterterms $\Delta_\sscript{UV}(S)$ of the subgraph $ S $ and the IR counterterms $\Delta_\sscript{IR}(S')$ of the subgraph $ S' $ are inserted back into the remnant graph $ G/S \setminus S'$ with the $ \ast$-operation, which for the UV subdivergences simply replaces the disjoint subgraphs with the Feynman rule for the counterterm. 
Ultimately, Eq.~\eqref{eq:BarRStarOperation} relates the UV counterterm $\Delta_\sscript{UV}(G)$ of the full graph $G$ to $ \Bar{\boldsymbol{R}}^{\ast}G $, which subtracts all UV and IR subdivergences from the graph, and the finite object $ \boldsymbol{R}^{\ast}G $.

Working in dimensional regularization, we introduce the $\boldsymbol{K}$-operation to extract the $\epsilon$-poles from an expression:
\begin{align}\label{eq:Koperation}
    \boldsymbol{K} \! \! \sum_{n= \eminus \infty}^\infty c_n \epsilon^n = \sum_{n=1}^\infty c_{\eminus n} \frac{1}{\epsilon^n} \, .
\end{align}
In the MS and, by extension, $\overline{\mathrm{MS}}$ schemes, counterterms consist entirely of divergent pieces.
Thus, by applying \eqref{eq:Koperation} to \eqref{eq:BarRStarOperation} and noting that $\boldsymbol{R}^{\ast} G$ is finite (by construction), we get
\begin{align}
    \Delta_\sscript{UV}(G) = - \boldsymbol{K} \Bar{\boldsymbol{R}}^{\ast} G \, .
\end{align}
It is well-known that the counterterms in the minimal subtraction scheme are homogeneous polynomials of mass-weighted degree $\omega(G)$ in the masses and external momenta, where $\omega(G)$ denotes the superficial degree of divergence of $G$. The Taylor expansion operator $\boldsymbol{T}^{\omega(G)}_{\{p_i, m_j^2\}}$, which returns the term of order $\omega(G)$ in the expansion around $p_i = m_j = 0$, therefore acts trivially on the UV counterterm: $ \boldsymbol{T}^{\omega(G)}_{\{p_i, m_j^2\}} \Delta_\sscript{UV}(G) = \Delta_\sscript{UV}(G)  $.
Moreover, the Taylor expansion operator commutes with the $\Bar{\boldsymbol{R}}^{\ast}$-operation, allowing us to write
\begin{align} \label{eq:ct_formula_Rstar}
    \Delta_\sscript{UV}(G) = - \boldsymbol{K} \Bar{\boldsymbol{R}}^{\ast} \boldsymbol{T}^{\omega(G)}_{\{p_i, m_j^2\}} G \, .
\end{align}
Even though the expansion around zero masses and external momenta introduces spurious IR divergences, the $ \Bar{\boldsymbol{R}}^{\ast} $-operation subtracts these from the result, and the right-hand side remains well-behaved.
The Taylor expanded graph $G$ is a logarithmically divergent vacuum graph, which has a UV divergence even though it is scaleless.  
It is well-established (see, e.g., the discussion in Ref.~\cite{Beekveldt:2020kzk}) that one may IR rearrange logarithmically divergent graphs without changing their UV counterterm; that is, one may change/introduce masses and external momenta. We can therefore introduce a common auxiliary mass in all propagators, thus regulating all IR divergences while reducing the integration to a single-scale integral. It follows that the part of the $\Bar{\boldsymbol{R}}^{\ast}$-operation responsible for subtracting IR divergences becomes trivial with this approach.

\subsubsection{Adaptation of the \texorpdfstring{$ R^{\ast} $}{R*}-method to functional methods}
So far, the discussion of the $\boldsymbol{R}^{\ast}$-method has been on the level of Feynman graphs. One can also make it work at the integrand level, although one should be careful about contracting Lorentz tensor structures because this can produce factors of $ d=4- 2\epsilon$, which do not commute with the $\boldsymbol{K}$-operator~\cite{Herzog:2017bjx}. Consequently, all contractions should be performed only after the counterterm formula~\eqref{eq:ct_formula_Rstar} has been applied recursively (with appropriate IR rearrangement). In this manner, we are sure that the UV counterterms inserted by the $ \Bar{\boldsymbol{R}}^{\ast} $-operator correctly reproduce the counterterms of the renormalized Lagrangian. 
With this caveat, it becomes possible to apply the counterterm formula directly to the saddlepoint approximation for the effective action~\eqref{eq:effective_action}. The counterterm action at two-loop order is then
    \begin{equation}\label{eq:TwoLoopCT}
    \widehat{S}^{(2)} = \dfrac{1}{8} \boldsymbol{K} \bar{\boldsymbol{R}}^\ast \boldsymbol{T} \Big[ \widehat{\mathcal{D}}^{(0)}_{IJKL} \widehat{\mathcal{Q}}_{IJ}^{\eminus1} \widehat{\mathcal{Q}}_{KL}^{\eminus1} \Big]
    - \dfrac{1}{12} \boldsymbol{K} \bar{\boldsymbol{R}}^\ast \boldsymbol{T} \Big[ \widehat{\mathcal{C}}_{IJK}^{(0)} \widehat{\mathcal{Q}}_{IL}^{\eminus1} \widehat{\mathcal{Q}}_{JM}^{\eminus1} \widehat{\mathcal{Q}}_{KN}^{\eminus1} \widehat{\mathcal{C}}^{(0)}_{LMN} \Big],
    \end{equation}
keeping in mind the delayed contraction of tensor structures in the loop integrands. The insertion of one-loop counterterms in the one-loop topology is absent in the formula, as it is automatically reproduced by the $\Bar{\boldsymbol{R}}^{\ast}$-operation on the genuine two-loop topologies.

To recap, the general recipe we use for evaluating the counterterm associated with any two-loop Feynman-integral $G$ in the background-field formalism is the following:
\begin{enumerate}
    \item \label{enum:taylor_expand} Taylor-expand $G$ in its masses, background fields, and covariant momenta operator and keep only the terms of mass-weighted order $\omega(G)$, where $\omega(G)$ is the superficial degree of divergence of $G$. 
    \item Introduce a single auxiliary mass $a$ to all propagators of the resulting logarithmically divergent scaleless integrals as an IR regulator.
    \item Apply the $\Bar{\boldsymbol{R}}^{\ast}$-operation to the now single scale vacuum integrals by
    \begin{enumerate}
        \item constructing all UV divergent subdiagrams $S$ and their corresponding remaining diagram $G/S$. Due to the regulator mass $a$, there are no IR divergences, so $S'$ is trivially the empty set.
        \item evaluating the counterterm $\Delta_\sscript{UV}(S)$ of each divergent subdiagram $S$, using the same procedure as for the full graph, i.e., starting again from step~\ref{enum:taylor_expand} but with $G=S$.\footnote{In the subdivergence the Taylor expansion treats the auxiliary mass $a$ and external loop momenta as small compared to the loop momenta of $S$.} 
    \end{enumerate}
    \item Evaluate the momentum integrals, which are now single-scale massive tadpoles, using, e.g., the integration formulas provided in~\cite{Chetyrkin:1997fm}.
    \end{enumerate}

The $\boldsymbol{R}^{\ast}$-method, as we have described it in this section, is very convenient for the functional approach to counterterm evaluation. First of all, the Taylor expansion gives a handle on the $ \mathcal{Q}^{\eminus 1} $ similarly to the hard-region expansion used in matching calculations. Another, less obvious, benefit is that the $ \Bar{\boldsymbol{R}}^{\ast}$-operator automatically subtracts subdivergences stemming from nuisance operators, such as operators proportional to the equations of motion, as described in~\cite{Collins:1994ee,Naterop:2023dek} and references therein. These are necessary to remove subdivergences and extract the overall counterterm but do not mix back into the physical operators under the RG. It is, therefore, convenient to avoid having to construct these counterterms explicitly.

\subsubsection{A sample application of the \texorpdfstring{$ R^{\ast} $}{R*}-method}
At two-loop order there are only two topologies: the figure-8 and the sunset topologies, of which the sunset is the only one that does not factor into a product of two one-loop topologies. In general, the sunset consists of three propagators $ P_{1,2,3} $ and two vertices $ V_{1,2} $ (which in the functional expression are functions of the background fields). Our approach is to always use IR rearrangement to introduce masses to all propagators, thereby removing all IR singularities, before applying the \rstaroperation to the loop graph. Thus, $ \boldsymbol{R}^{\ast} $ does not introduce any IR counterterms and becomes equivalent to the ordinary $ \boldsymbol{R} $-operation. For the sunset, we schematically have
    \begin{equation} \label{eq:Rop_on_sunset}
    \bar{\boldsymbol{R}}^{\ast}
    \ineqgraphics{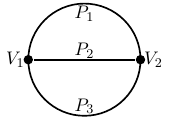}
    = \; \ineqgraphics{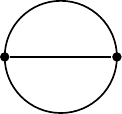}
    \, +\,  
    \ineqgraphics{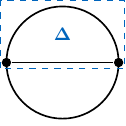}
    \, +\,
        \ineqgraphics{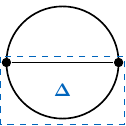}  
    \, +\,
    \ineqgraphics{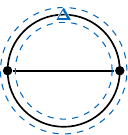}
    \, .
    \end{equation}
The subgraph enclosed in dashed blue lines in the last three terms implies the insertion of a UV counterterm in place of the subgraph. 

We will now demonstrate how to use the \rstaroperation to obtain the counterterms for a loop graph. In particular, we consider the expression
    \begin{equation}
    I= \int_{k\ell q} \! \delta(k+\ell +q) \left[ g_{\mu\nu} \right]_{V_1} \ast \left[ \dfrac{g^{\nu\rho}}{k^2- m^2} \right]_{P_1} \!\! \ast \left[ \dfrac{1}{\ell^2} \right]_{P_2} \!\! \ast \left[ \dfrac{g^{\mu\sigma}}{q^2} \right]_{P_3} \!\! \ast \left[ g_{\rho\sigma} \right]_{V_2} .
    \end{equation}
This is not quite an ordinary loop integral: at this stage we need to carefully discriminate the origin of all terms, that is, which propagator ($ P_i $) or vertex ($ V_i $) they stem from. The notation `$ \ast $' is used to indicate eventual multiplication of the various terms, but only after application of the \rstaroperation. That is because $ \boldsymbol{R}^{\ast} $ replaces pieces of the two-loop integral with the corresponding counterterms, which is possible only if we keep track of their origin. For the same reason, we are not allowed to contract any Lorentz indices between the various propagator and vertex terms before the \rstaroperation is applied.

To determine the UV counterterm associated with $ I $, we begin by applying the Taylor expansion, which expands the integral in physical scales and selects the logarithmically divergent pieces:
    \begin{equation}
    \begin{split}
    \ctop I &= - \kop \bar{\boldsymbol{R}}^\ast \boldsymbol{T} I \\
    &= -m^2 \kop \bar{\boldsymbol{R}}^\ast \! \int_{k\ell q} \! \delta(k+\ell +q) \left[ g_{\mu\nu} \right]_{V_1} \ast \left[ \dfrac{g^{\nu\rho}}{k^4} \right]_{P_1} \!\! \ast \left[ \dfrac{1}{\ell^2} \right]_{P_2} \!\! \ast \left[ \dfrac{g^{\mu\sigma}}{q^2} \right]_{P_3} \!\! \ast \left[ g_{\rho\sigma} \right]_{V_2}.
    \end{split}
    \end{equation}
The factor $ m^2 $ can be extracted from the integral, as it is an overall factor from the perspective of $ \boldsymbol{R}^{\ast} $ and it commutes with $ \kop $, since it does not depend on $ \epsilon $ in any way. The Taylor expansion leaves us with a scaleless integral, but even so, it does not vanish because of the \rstaroperation. To proceed, we are free to apply IR rearrangement to our logarithmically divergent integral. We simply add an auxiliary mass (squared) parameter $ a $ to all propagators and obtain 
    \begin{equation}
    \ctop I = -m^2 \kop \bar{\boldsymbol{R}}^\ast \! \int_{k\ell q} \! \delta(k+\ell +q) \left[ g_{\mu\nu} \right]_{V_1} \ast \left[ \dfrac{g^{\nu\rho}}{(k^2 -a)^2} \right]_{P_1} \!\! \ast \left[ \dfrac{1}{\ell^2- a} \right]_{P_2} \!\! \ast \left[ \dfrac{g^{\mu\sigma}}{q^2- a} \right]_{P_3} \!\! \ast \left[ g_{\rho\sigma} \right]_{V_2}.
    \end{equation}
Now that the integral is IR safe, $ \bar{\boldsymbol{R}}^{\ast} $ acts as in Eq.~\eqref{eq:Rop_on_sunset}. We find 
    \begin{equation} \label{eq:metric_sunset_example}
    \begin{split}
    \ctop I &= -m^2 \kop \bigg( d\! \int_{k\ell } \, \dfrac{1}{(k^2-a)^2} \dfrac{1}{\ell^2-a} \dfrac{1}{(k+ \ell)^2 -a} \\ 
    &\qquad + \int_k\, \left[ \dfrac{g^{\nu\rho}}{(k^2-a)^2} \right]_{P_1} \!\! \ast \ctop \!\! \int_{\ell q}\! \delta(k+\ell +q) \left[ g_{\mu\nu} \right]_{V_1} \ast \left[ \dfrac{1}{\ell^2- a} \right]_{P_2} \!\! \ast \left[ \dfrac{g^{\mu\sigma}}{q^2- a} \right]_{P_3} \!\! \ast \left[ g_{\rho\sigma} \right]_{V_2} \bigg).
    \end{split}
    \end{equation}
In the first term, without any counterterm insertions, it is no longer necessary to keep the factors from different vertices and propagators separate, and they simply multiply. Of the three one-loop subgraphs, only one of them is divergent. The UV counterterms of the other two subgraphs (with propagators $ P_1, P_2 $ and $ P_1, P_3 $) are zero, as they are superficially convergent. 

The counterterm for the one-loop subdivergence of $ I $ may now be evaluated. Here the mass parameter $ a $ and the loop momenta $ k $ behave as external parameters. The Taylor-expansion operator $ \boldsymbol{T} $ will expand in these parameters, but since the graph is already logarithmically divergent, its action is setting $ a=k_\mu=0 $.\footnote{The reader may be tempted to keep $ a $ as in IR regulator rather than setting it to $ 0 $ as part of the Taylor expansion. We consider it conceptually cleaner to fully execute the $ \boldsymbol{T} $ operation and then introduce a new mass regulator to emphasize that $ a $ is part of the two-loop IR rearrangement and external to the counterterms of the resulting subdivergences.} IR rearrangement lets us introduce another mass (squared) parameter $ b $ to IR regulate the resulting graph. At that point, the $ \bar{\boldsymbol{R}}^{\ast} $ operator acts trivially. The whole expression evaluates, as 
    \begin{equation}
    \ctop \!\! \int_{\ell q}\! \delta(k+\ell +q) \left[ g_{\mu\nu} \right]_{V_1} \ast \left[ \dfrac{1}{\ell^2- a} \right]_{P_2} \!\! \ast \left[ \dfrac{g^{\mu\sigma}}{q^2- a} \right]_{P_3} \!\! \ast \left[ g_{\rho\sigma} \right]_{V_2} 
    = - \kop \! \left(g^{\nu\rho} \! \int_\ell \; \dfrac{1}{(\ell^2 - b)^2} \right) 
    = \dfrac{1}{ 16 \pi^2} \dfrac{ i}{\epsilon} g^{\nu\rho}.
    \end{equation}
Note that the UV counterterm for the subgraph does not depend on the spurious $ b $ parameter that we introduced as an IR regulator. 

Having evaluated the counterterm of the one-loop subdivergence, we may proceed with the counterterm evaluation for $ I $. There is no more diagram structure to keep track of and the last `$ \ast $' is replaced with ordinary multiplication:
    \begin{equation}
    \begin{split}
    \ctop I &= -m^2 \kop \bigg( d\! \int_{k\ell} \,\dfrac{1}{(k^2-a)^2} \dfrac{1}{\ell^2-a} \dfrac{1}{(k+\ell)^2-a} + \dfrac{1}{16 \pi^2} \dfrac{i}{\epsilon}\, d\! \int_k\, \dfrac{1}{(k^2-a)^2} \bigg) \\
    &= - \dfrac{m^2}{(16 \pi^2)^2}  \left( \dfrac{2}{\epsilon^2} - \dfrac{3}{\epsilon} \right).
    \end{split}
    \end{equation}
The resulting counterterm for the two-loop integral is polynomial in the model parameters (the mass). Crucially, there is no dependence on the spurious mass parameter $ a $ (although the two terms in \eqref{eq:metric_sunset_example} separately depend logarithmically on $ a $). This agrees with the assumption of the UV counterterm being independent of IR rearrangement.

\subsection{Combinatoric considerations} \label{subsec:combinatorics}
Following Section 2.4 in Ref. \cite{Fuentes-Martin:2023ljp}, we decompose the dressed propagators as 
\begin{align}
    Q^{\eminus 1}_{ab}(x,P_x+k) = \lzm \frac{1}{\Delta^{\eminus 1}(P_x + k)-X(x, P_x + k)} \dzm_{ab},
\end{align}
with the dressed momentum-space propagator $\Delta_{ab}$ and the dressed two-point interaction terms $X_{ab}$. 
Of these terms, $ \Delta^{\eminus 1}$ is dominant at large loop momenta, and the dressed propagator is expanded as 
\begin{equation}\label{eq:OpProdExp}
    Q^{\eminus 1}_{ab}(x,P_x+k) =  \sum^\infty_{n=0} (\Delta X)^n_{ac} \, \Delta_{cb},
\end{equation}
where the sum over $n$ is usually truncated at the required EFT order.

To reduce the total number of terms that appear in the sunset and figure-8 topologies, we examine the structure of \eqref{eq:OpProdExp} in light of the internal symmetries of the functional contraction.\footnote{This is a lot like accounting for symmetries in ordinary Feynman diagram to obtain the right symmetry factor while avoiding evaluating identical diagrams multiple times.} We focus on the types of propagators ($S\equiv$ Scalar, $V\equiv$ Vector, etc.) that are generated in this expansion and leave out the $X_{ab}$ terms for now (they are uniquely determined by the propagator types) and schematically write down, e.g.,
\begin{align}
    \begin{split}
    Q^{-1}_{SS}&\sim \Delta_{S}+\Delta_{S}\Delta_{S}+\Delta_{S}\Delta_{S}\Delta_{S}+\Delta_{S}\Delta_{V}\Delta_{S}+\ldots\\
    &=\{\text{Scalar}\}+\{\text{Scalar},\text{Scalar}\}+\{\text{Scalar},\text{Scalar},\text{Scalar}\}\\
    &+\{\text{Scalar},\text{Vector},\text{Scalar}\}+\ldots\,,
    \end{split}\\
    \begin{split}
    Q^{-1}_{SV}&\sim \Delta_{S}\Delta_{V}+\Delta_{S}\Delta_{S}\Delta_{V}+\Delta_{S}\Delta_{V}\Delta_{V}+\ldots\\
    &=\{\text{Scalar},\text{Vector}\}+\{\text{Scalar},\text{Scalar},\text{Vector}\}+\{\text{Scalar},\text{Vector},\text{Vector}\}+\ldots\,,
    \end{split}
\end{align}
and similarly for $	Q^{-1}_{VS}$ and $	Q^{-1}_{VV}$. Cases involving ghosts are obtained similarly. 

Next, we consider the relevant types of two-loop topologies---namely, the sunset and the figure-8. The expression for sunset type topology, $G_{\mathrm{ss.}}$, contains a triplet of dressed propagators $\{Q^{-1},Q^{-1},Q^{-1}\}$, whereas the figure-8 type topology, $G_{\mathrm{f8.}}$, can be written in terms of  doublets $\{Q^{-1},Q^{-1}\}$. Once the $ \Delta $'s have been specified, the vertices follow uniquely, just as the $ X $ terms do. In each topology, there is a symmetry associated with the exchange of dressed propagators. In the sunset there is an additional symmetry from exchanging the vertices, or, equivalently, from simultaneous reflection of all three propagators. The figure-8 is symmetric under the reflection of each propagator individually. Performing the expansion above and collecting the unique terms, we obtain
\begin{align}
    \begin{split}
    \{Q^{-1}_{SS},Q^{-1}_{SS},Q^{-1}_{SV}\}&\sim6\times\{\{\text{Scalar}\},\{\text{Scalar}\},\{\text{Scalar},\text{Vector}\}\}\\
    &+12\times\{\{\text{Scalar},\text{Scalar}\},\{\text{Scalar}\},\{\text{Scalar},\text{Vector}\}\}+\ldots,
    \end{split}\\
    \begin{split}
    \{Q^{-1}_{SS},Q^{-1}_{SV}\}&\sim4\times\{\{\text{Scalar}\},\{\text{Scalar},\text{Vector}\}\}\\
    &+4\times\{\{\text{Scalar},\text{Scalar}\},\{\text{Scalar},\text{Vector}\}\}+\ldots,
    \end{split}
\end{align}
and similarly for all other choices of propagators. Each prefactor here is basically a symmetry factor associated with the type of topology. For sunset-type topology possible prefactors are 1, 2, 3, 6, and 12, while for the figure-8 they are 1, 2, 4, and 8. This procedure significantly reduces the number of terms that have to be computed.

Further symmetrizations can be performed when specifying the possible insertions of field indices in $X_{ab}$ and the vertices $C_{abc}$ and $D_{abcd}$. For example, in a model that contains a complex scalar $\phi$ there are two possible substitutions for $\{\text{Scalar}\}$, namely $\{\phi\}$ and $\{\phi^\ast\}$. This means that for a topology of type  $\{\{\text{Scalar}\},\{\text{Scalar}\},\{\text{Scalar}\}\}$ with symmetry factor 1 there are three ways to have 2$\times\{\phi\}$  and $1\times\{\phi^\ast\}$ insertions. Similar consideration holds for all other types of terms and field insertions. 
Thus, the field symmetry factor, $s_{\text{f}}$, comprises all possible permutations and reflections of fields. The two symmetry factors $ s_\mathrm{t}$ associated with the propagator types and $ s_\mathrm{f}$ arising from the particular field insertions can be multiplied together to obtain the total symmetry factor 
\begin{equation}
    s_{\text{tf}}=s_{\text{t}}\times s_{\text{f}}.
\end{equation}

Final symmetries arise in the sums over EFT orders in $G_{\mathrm{ss.}}$. Here the symmetry of the expression in the last two dressed propagators can be exploited and an additional symmetry factor, $s_{\text{o}}$, obtained. This gives the total symmetry factor $s_{\text{tfo}}$ associated with the type, field insertion, and order expansion:
\begin{equation}
	s_{\text{tfo}}=s_{\text{t}}\times s_{\text{f}}\times s_{\text{o}},
\end{equation}
which can be factored out of every unique term computed in $G_{\mathrm{ss.}}$. Figure-8 topologies are less computationally demanding. Hence, we consider only the $s_{\text{tf}}$ symmetry factors in the evaluation of $G_{\mathrm{f8.}}$.

\subsection{Practical example}
We now demonstrate the functional techniques with a minimal example: a massless real scalar field $\Phi$ with a $Z_2$ symmetry. The Lagrangian of the theory is given by 
\begin{align}
    \mathcal{L}(\Phi) = \frac{1}{2} \partial_\mu \Phi \, \partial^\mu \Phi - \frac{\lambda}{4!} \Phi^4 \, .
\end{align}
The field $\Phi$ is split into a quantum field $\phi$ and a classical background field $\hat{\phi}$\,: 
\begin{align}
    \Phi \rightarrow \phi + \Hat{\phi} \, .
\end{align}
As a first step, we derive all X-insertions as well as the vertices to the two-loop topologies from the expansion of the action~\eqref{eq:action_expansion}:
\begin{align}
    X_{\phi \phi}(\Hat{\phi}) = \frac{\lambda}{2}\Hat{\phi}^2 \,, \qquad C_{\phi \phi \phi}(\Hat{\phi}) = -\lambda \Hat{\phi}\,, \qquad D_{\phi\phi\phi\phi}(\Hat{\phi}) = - \lambda \, .
\end{align}
\begin{figure}
    \centering
    \def\radius{0.7} 
    \begin{subfigure}{0.24\textwidth}
        \centering
        \begin{tikzpicture}
            \draw[] (0,0) circle [radius=\radius];
            \draw[] (-\radius,0) -- (\radius,0);
            \filldraw[black] (-\radius,0) circle (2pt); 
            \filldraw[black] (\radius,0) circle (2pt);
            \draw[dashed, color=blue] (-\radius-0.07,0) -- (-2*\radius,0);
            \draw[dashed, color=blue] (\radius+0.07,0) -- (2*\radius,0);
        \end{tikzpicture}
        \caption{}
        \label{subfig:Sunset1}
    \end{subfigure}
    \begin{subfigure}{0.24\textwidth}
        \centering
        \begin{tikzpicture}
            \draw[] (0,0) circle [radius=\radius];
            \draw[] (-\radius,0) -- (\radius,0);
            \filldraw[black] (-\radius,0) circle (2pt); 
            \filldraw[black] (\radius,0) circle (2pt);
            \filldraw[black] (0,\radius) circle (2pt);
            \draw[dashed, color=blue] (-\radius-0.07,0) -- (-2*\radius,0);
            \draw[dashed, color=blue] (\radius+0.07,0) -- (2*\radius,0);
            \draw[dashed, color=blue] (-0.05,\radius+0.05) -- (-0.5*\radius,1.5*\radius);
            \draw[dashed, color=blue] (0.05,\radius+0.05) -- (0.5*\radius,1.5*\radius);
        \end{tikzpicture}
        \caption{}
        \label{subfig:Sunset2}
    \end{subfigure}
     \begin{subfigure}{0.24\textwidth}
        \centering
        \begin{tikzpicture}
            \draw[] (0,0) circle [radius=\radius];
            \draw[] (-\radius,0) -- (\radius,0);
            \filldraw[black] (-\radius,0) circle (2pt); 
            \filldraw[black] (\radius,0) circle (2pt);
            \filldraw[black] (0,0) circle (2pt);
            \draw[dashed, color=blue] (-\radius-0.07,0) -- (-2*\radius,0);
            \draw[dashed, color=blue] (\radius+0.07,0) -- (2*\radius,0);
            \draw[dashed, color=blue] (-0.05,0.05) -- (-0.5*\radius,0.5*\radius);
            \draw[dashed, color=blue] (0.05,0.05) -- (0.5*\radius,0.5*\radius);
        \end{tikzpicture}
        \caption{}
        \label{subfig:Sunset3}
    \end{subfigure}
     \begin{subfigure}{0.24\textwidth}
        \centering
        \begin{tikzpicture}
            \draw[] (0,0) circle [radius=\radius];
            \draw[] (-\radius,0) -- (\radius,0);
            \filldraw[black] (-\radius,0) circle (2pt); 
            \filldraw[black] (\radius,0) circle (2pt);
            \filldraw[black] (0,-\radius) circle (2pt);
            \draw[dashed, color=blue] (-\radius-0.07,0) -- (-2*\radius,0);
            \draw[dashed, color=blue] (\radius+0.07,0) -- (2*\radius,0);
            \draw[dashed, color=blue] (-0.05,-\radius+0.05) -- (-0.5*\radius,-0.5*\radius);
            \draw[dashed, color=blue] (0.05,-\radius+0.05) -- (0.5*\radius,-0.5*\radius);
        \end{tikzpicture}
        \caption{}
        \label{subfig:Sunset4}
    \end{subfigure}
    
    \vspace{0.3cm}

    \def\radius{0.5} 
    \begin{subfigure}{0.24\textwidth}
        \centering
        \begin{tikzpicture}
            \draw[] (-\radius,0) circle (\radius);
            \draw[] (\radius,0) circle (\radius);
            \filldraw[black] (0,0) circle (2pt); 
        \end{tikzpicture}
        \caption{}
        \label{subfig:Fig8_1}
    \end{subfigure}
    \begin{subfigure}{0.24\textwidth}
        \centering
        \begin{tikzpicture}
            \draw[] (-\radius,0) circle (\radius);
            \draw[] (\radius,0) circle (\radius);
            \filldraw[black] (0,0) circle (2pt); 
            \filldraw[black] (-2*\radius,0) circle (2pt); 
            \draw[dashed, color=blue] (-2*\radius-0.05,0.05) -- (-2.8*\radius,0.8*\radius);
            \draw[dashed, color=blue] (-2*\radius-0.05,-0.05) -- (-2.8*\radius,-0.8*\radius);
        \end{tikzpicture}
        \caption{}
        \label{subfig:Fig8_2}
    \end{subfigure}
    \begin{subfigure}{0.24\textwidth}
        \centering
        \begin{tikzpicture}
            \draw[] (-\radius,0) circle (\radius);
            \draw[] (\radius,0) circle (\radius);
            \filldraw[black] (0,0) circle (2pt); 
            \filldraw[black] (2*\radius,0) circle (2pt); 
            \draw[dashed, color=blue] (2*\radius+0.05,0.05) -- (2.8*\radius,0.8*\radius);
            \draw[dashed, color=blue] (2*\radius+0.05,-0.05) -- (2.8*\radius,-0.8*\radius);
        \end{tikzpicture}
        \caption{}
        \label{subfig:Fig8_3}
    \end{subfigure}
    \begin{subfigure}{0.24\textwidth}
        \centering
        \begin{tikzpicture}
            \draw[] (-\radius,0) circle (\radius);
            \draw[] (\radius,0) circle (\radius);
            \filldraw[black] (0,0) circle (2pt); 
            \filldraw[black] (-2*\radius,0) circle (2pt); 
            \filldraw[black] (2*\radius,0) circle (2pt); 
            \draw[dashed, color=blue] (-2*\radius-0.05,0.05) -- (-2.8*\radius,0.8*\radius);
            \draw[dashed, color=blue] (-2*\radius-0.05,-0.05) -- (-2.8*\radius,-0.8*\radius);
            \draw[dashed, color=blue] (2*\radius+0.05,0.05) -- (2.8*\radius,0.8*\radius);
            \draw[dashed, color=blue] (2*\radius+0.05,-0.05) -- (2.8*\radius,-0.8*\radius);
        \end{tikzpicture}
        \caption{}
        \label{subfig:Fig8_4}
    \end{subfigure}
    
    \vspace{0.3cm}

     \begin{subfigure}{0.24\textwidth}
        \centering
        \begin{tikzpicture}
            \draw[] (-\radius,0) circle (\radius);
            \draw[] (\radius,0) circle (\radius);
            \filldraw[black] (0,0) circle (2pt); 
            \filldraw[black] (-\radius-\radius/1.41,\radius/1.41) circle (2pt); 
            \filldraw[black] (-\radius-\radius/1.41,-\radius/1.41) circle (2pt);
            \draw[dashed, color=blue] (-\radius-\radius/1.41,\radius/1.41+0.05) -- (-\radius-\radius/1.41,\radius/1.41+0.8*\radius);
            \draw[dashed, color=blue] (-\radius-\radius/1.41-0.05,\radius/1.41) -- (-\radius-\radius/1.41-0.8*\radius,\radius/1.41);
            \draw[dashed, color=blue] (-\radius-\radius/1.41,-\radius/1.41-0.05) -- (-\radius-\radius/1.41,-\radius/1.41-0-0.8*\radius);
            \draw[dashed, color=blue] (-\radius-\radius/1.41-0.05,-\radius/1.41) -- (-\radius-\radius/1.41-0.8*\radius,-\radius/1.41);
        \end{tikzpicture}
        \caption{}
        \label{subfig:Fig8_5}
    \end{subfigure}
     \begin{subfigure}{0.24\textwidth}
        \centering
        \begin{tikzpicture}
            \draw[] (-\radius,0) circle (\radius);
            \draw[] (\radius,0) circle (\radius);
            \filldraw[black] (0,0) circle (2pt); 
            \filldraw[black] (\radius+\radius/1.41,\radius/1.41) circle (2pt); 
            \filldraw[black] (\radius+\radius/1.41,-\radius/1.41) circle (2pt);
            \draw[dashed, color=blue] (\radius+\radius/1.41,\radius/1.41+0.05) -- (\radius+\radius/1.41,\radius/1.41+0.8*\radius);
            \draw[dashed, color=blue] (\radius+\radius/1.41+0.05,\radius/1.41) -- (\radius+\radius/1.41+0.8*\radius,\radius/1.41);
            \draw[dashed, color=blue] (\radius+\radius/1.41,-\radius/1.41-0.05) -- (\radius+\radius/1.41,-\radius/1.41-0-0.8*\radius);
            \draw[dashed, color=blue] (\radius+\radius/1.41+0.05,-\radius/1.41) -- (\radius+\radius/1.41+0.8*\radius,-\radius/1.41);
        \end{tikzpicture}
        \caption{}
        \label{subfig:Fig8_6}
    \end{subfigure}
    \caption{All superficially divergent diagrams in the minimal example. The dashed blue lines denote external background fields, while the solid black lines are internal quantum fields. The black dots on propagator lines represent insertions of X-terms through the expansion around large loop momenta.}
    \label{fig:div_diags}
\end{figure}
Next, we construct all superficially divergent diagrams by expanding the dressed propagators $\mathcal{Q}^{\eminus 1}$ in \eqref{eq:TwoLoopCT} around large loop momenta, as given in \eqref{eq:OpProdExp}. The resulting diagrams are shown in Figure \ref{fig:div_diags}. Graphically, the propagators $\Delta$ are represented as solid lines, while all vertices as well as insertions of $X$-terms are denoted by dots. External background fields are shown with dashed blue lines.

We can immediately reduce the total number of diagrams from ten to seven by following the symmetry considerations given in Section \ref{subsec:combinatorics}: The diagrams \ref{subfig:Sunset2} through \ref{subfig:Sunset4} are related by exchanging propagator lines, which allows us to include only one of them and multiply it by a symmetry factor of 3. A similar argument holds for the diagrams \ref{subfig:Fig8_5} and \ref{subfig:Fig8_6}. The diagrams with positive superficial degree of divergence are then brought into logarithmically divergent form either by the expansion of $\Delta$ in \eqref{eq:OpProdExp} or via the Taylor expansion in \eqref{eq:Gss}.

As discussed earlier in Section~\ref{sec:RStar}, we apply the $\boldsymbol{R}^\ast$-method to extract the local counterterm associated with any of the two-loop graphs. In the case of diagram \ref{subfig:Sunset2}, we represent this procedure graphically as
\begin{equation}
    \begin{split}
    \Delta_\sscript{UV}(G_\text{ss.}) 
    &\supset - \frac{3}{12} \boldsymbol{K} \Bar{\boldsymbol{R}}^{\ast} 
    \left(  
        \def\radius{0.48} 
        \begin{tikzpicture}[baseline=-0.1cm]
            \draw[] (0,0) circle [radius=\radius];
            \draw[] (-\radius,0) -- (\radius,0);
            \filldraw[black] (-\radius,0) circle (2pt); 
            \filldraw[black] (\radius,0) circle (2pt);
            \filldraw[black] (0,\radius) circle (2pt);
            \draw[dashed, color=blue] (-\radius-0.07,0) -- (-2*\radius,0);
            \draw[dashed, color=blue] (\radius+0.07,0) -- (2*\radius,0);
            \draw[dashed, color=blue] (-0.05,\radius+0.05) -- (-0.5*\radius,1.5*\radius);
            \draw[dashed, color=blue] (0.05,\radius+0.05) -- (0.5*\radius,1.5*\radius);
        \end{tikzpicture} 
    \right) \\
    &=
    - \frac{3}{12}\boldsymbol{K}
    \left(
        \def\radius{0.48} 
        \begin{tikzpicture}[baseline=-0.1cm]
            \draw[] (0,0) circle [radius=\radius];
            \draw[] (-\radius,0) -- (\radius,0);
            \filldraw[black] (-\radius,0) circle (2pt); 
            \filldraw[black] (\radius,0) circle (2pt);
            \filldraw[black] (0,\radius) circle (2pt);
            \draw[dashed, color=blue] (-\radius-0.07,0) -- (-2*\radius,0);
            \draw[dashed, color=blue] (\radius+0.07,0) -- (2*\radius,0);
            \draw[dashed, color=blue] (-0.05,\radius+0.05) -- (-0.5*\radius,1.5*\radius);
            \draw[dashed, color=blue] (0.05,\radius+0.05) -- (0.5*\radius,1.5*\radius);
        \end{tikzpicture} 
    +
    \Delta_\sscript{UV}
    \left(
        \def\radius{0.48} 
        \begin{tikzpicture}[baseline=-0.1cm, rotate=180]
            \clip[] (-2*\radius,-0.2) rectangle (2*\radius,\radius);
            \draw[] (0,0) circle [radius=\radius];
            \draw[] (-\radius,0) -- (\radius,0);
            \filldraw[black] (-\radius,0) circle (2pt); 
            \filldraw[black] (\radius,0) circle (2pt);
            \draw[dashed, color=blue] (-\radius-0.07,0) -- (-2*\radius,0);
            \draw[dashed, color=blue] (\radius+0.07,0) -- (2*\radius,0);
        \end{tikzpicture}         
    \right)
    \ast
    \def\radius{0.48} 
    \begin{tikzpicture}[baseline=-0.1cm]
        \draw[] (0,0) circle [radius=\radius];
        \filldraw[black] (0,\radius) circle (2pt);
        \node[fill=white, inner sep=-.5pt] at (0,-\radius) {$\boxtimes$};
        \draw[dashed, color=blue] (-0.05,\radius+0.05) -- (-0.5*\radius,1.5*\radius);
        \draw[dashed, color=blue] (0.05,\radius+0.05) -- (0.5*\radius,1.5*\radius);
    \end{tikzpicture}  
    \right)
    \, ,
    \end{split}
\end{equation}
where $\boxtimes$ denotes the insertion point of the one-loop counterterm. The factor $\frac{1}{12}$ is the generic sunset prefactor of~\eqref{eq:TwoLoopCT}, while the factor 3 represents the symmetry of having the X-insertion on any of the three propagator lines. 
The first term on the right-hand side extracts the poles of the full diagram,
\begin{align}\label{eq:example_full_diagram}
\begin{split}
    - \frac{3}{12}\boldsymbol{K}
    \left(
        \def\radius{0.48} 
        \begin{tikzpicture}[baseline=-0.1cm]
            \draw[] (0,0) circle [radius=\radius];
            \draw[] (-\radius,0) -- (\radius,0);
            \filldraw[black] (-\radius,0) circle (2pt); 
            \filldraw[black] (\radius,0) circle (2pt);
            \filldraw[black] (0,\radius) circle (2pt);
            \draw[dashed, color=blue] (-\radius-0.07,0) -- (-2*\radius,0);
            \draw[dashed, color=blue] (\radius+0.07,0) -- (2*\radius,0);
            \draw[dashed, color=blue] (-0.05,\radius+0.05) -- (-0.5*\radius,1.5*\radius);
            \draw[dashed, color=blue] (0.05,\radius+0.05) -- (0.5*\radius,1.5*\radius);
        \end{tikzpicture} 
    \right)
    &= 
    - \frac{3}{12}\boldsymbol{K}
    \left(
        \frac{\lambda^3}{2} \Hat{\phi}^4 \int_{k\ell} \,\frac{1}{(k^2-a)^2 (\ell^2-a)[(k+ \ell)^2-a]}
    \right) \\
    &=
    \frac{\lambda^3}{16} \Hat{\phi}^4
    \left[
        \frac{1}{\epsilon^2} + \frac{1}{\epsilon} + \frac{2}{\epsilon} \log \! \left(\frac{\mu^2}{a}\right)
    \right] \, ,
\end{split}
\end{align}
where the auxiliary mass squared parameter $a$ was introduced to IR regulate the logarithmically divergent integral.
The second term on the right-hand side involves the computation of a one-loop subdivergence:
\begin{align}
    \Delta_\sscript{UV}
    \left(
        \def\radius{0.48} 
        \begin{tikzpicture}[baseline=-0.1cm, rotate=180]
            \clip[] (-2*\radius,-0.2) rectangle (2*\radius,\radius);
            \draw[] (0,0) circle [radius=\radius];
            \draw[] (-\radius,0) -- (\radius,0);
            \filldraw[black] (-\radius,0) circle (2pt); 
            \filldraw[black] (\radius,0) circle (2pt);
            \draw[dashed, color=blue] (-\radius-0.07,0) -- (-2*\radius,0);
            \draw[dashed, color=blue] (\radius+0.07,0) -- (2*\radius,0);
        \end{tikzpicture}         
    \right)
    =
    - \boldsymbol{K} \Bar{\boldsymbol{R}}^{\ast} \boldsymbol{T} 
    \left(
        \lambda^2 \Hat{\phi}^2 \int_{\ell} \,\frac{1}{(\ell^2-a) [(k+\ell)^2-a]}
    \right) \, .
\end{align}
The operator $\boldsymbol{T}$ expands the integrand around large values of $\ell$, keeping only logarithmically divergent terms as everything else vanishes in dimensional regularization. In our simple example, this amounts to setting $a=k_\mu=0$. $ \boldsymbol{R}^\ast$ now acts trivially on the integrand since there is no further subdivergence to subtract. The resulting integral, which is rendered IR safe by introducing another auxiliary mass squared parameter $b$, then evaluates to  
\begin{align}
    \Delta_\sscript{UV}
    \left(
        \def\radius{0.48} 
        \begin{tikzpicture}[baseline=-0.1cm, rotate=180]
            \clip[] (-2*\radius,-0.2) rectangle (2*\radius,\radius);
            \draw[] (0,0) circle [radius=\radius];
            \draw[] (-\radius,0) -- (\radius,0);
            \filldraw[black] (-\radius,0) circle (2pt); 
            \filldraw[black] (\radius,0) circle (2pt);
            \draw[dashed, color=blue] (-\radius-0.07,0) -- (-2*\radius,0);
            \draw[dashed, color=blue] (\radius+0.07,0) -- (2*\radius,0);
        \end{tikzpicture}         
    \right)
    =
    - \boldsymbol{K}  
    \left(
        \lambda^2 \Hat{\phi}^2 \int_{\ell} \, \frac{1}{(\ell^2-b)^2}
    \right)
    = 
    -\frac{i \lambda^2}{16\pi^2 \epsilon} \Hat{\phi}^2
    \equiv
    G_\sscript{ct} \, .
\end{align}
This one-loop counterterm is now inserted into the remaining diagram:
\begin{align}
\begin{split}
    - \frac{3}{12}\boldsymbol{K}
    \left(
    G_\sscript{ct}
    \ast
    \def\radius{0.48} 
    \begin{tikzpicture}[baseline=-0.1cm]
        \draw[] (0,0) circle [radius=\radius];
        \filldraw[black] (0,\radius) circle (2pt);
        \node[fill=white, inner sep=-.5pt] at (0,-\radius) {$\boxtimes$};
        \draw[dashed, color=blue] (-0.05,\radius+0.05) -- (-0.5*\radius,1.5*\radius);
        \draw[dashed, color=blue] (0.05,\radius+0.05) -- (0.5*\radius,1.5*\radius);
    \end{tikzpicture}  
    \right)
    &=
    - \frac{3}{12}\boldsymbol{K}
    \left(
        \left[ -\frac{i \lambda^2}{16\pi^2 \epsilon} \Hat{\phi}^2 \right] \frac{\lambda}{2}\Hat{\phi}^2 \int_k \, \frac{1}{(k^2-a)^2}
    \right) \\
    &=
    -\frac{\lambda^3}{8(16\pi^2)^2} \Hat{\phi}^4
    \left[
        \frac{1}{\epsilon^2} + \frac{1}{\epsilon} \log \! \left(\frac{\mu^2}{a}\right)
    \right] \, .
\end{split}
\end{align}
By adding the result to \eqref{eq:example_full_diagram}, we obtain
\begin{align}
    \Delta_\sscript{UV}(G_\text{ss.}) \supset \frac{\lambda^3}{16(16\pi^2)^2} \Hat{\phi}^4
    \left(
        -\frac{1}{\epsilon^2} + \frac{1}{\epsilon} 
    \right)
\end{align}
for the local counterterm generated by the sum of diagrams \ref{subfig:Sunset2} through \ref{subfig:Sunset4}. Repeating this procedure for the remaining diagrams in \ref{fig:div_diags} would let us determine the full two-loop counterterm action $S^{(2)}$.

\section{Implementation and Results} \label{sec:results}
We now turn to a semi-realistic application of our methods for determining the UV counterterms, and thereby \befs. Their implementation in a computer code lets us determine the \befs of the bosonic SMEFT to two-loop order. While the addition of fermions is required in practical phenomenological studies, our results demonstrate the effectiveness of the methods for realistic models while providing an important benchmark against which to compare future results.

\subsection{Bosonic SMEFT}
The Lagrangian of the bosonic SMEFT is given by five pieces, namely the bosonic part of the SM, a total of 15 dimension-six operators, the gauge-fixing and ghost terms and the topological $\theta$ terms:
\begin{align}
    \L = \L_\sscript{SM} + \sum_{i=1}^{15} C_i \cO_i  + \L_\mathrm{gf.} + \L_\mathrm{gh.} + \L_\theta \, .
\end{align}
The bosonic SM contains the usual gauge fields associated with the full unbroken SM gauge group as well as the Higgs field $H$:
\begin{align}
    \L_\sscript{SM} = -\frac{1}{4} B^{\mu \nu} B_{\mu \nu} - \frac{1}{4} W^{I \, \mu \nu} W_{\mu \nu}^I - \frac{1}{4} G^{A \, \mu \nu} G_{\mu \nu}^A + D_\mu H^\dag D^\mu H + \mu^2 H^\dag H - \frac{\lambda}{2} (H^\dag H)^2 \, ,
\end{align}
where the gauge couplings appear in the covariant derivative as
\begin{align}
    D_\mu = \partial_\mu - i g_Y Y B_\mu - i g_L t^I W_\mu^I - i g_s t^A G_\mu^A \, ,
\end{align}
with $t^I = \frac{1}{2} \sigma^I$ being the generators of $\SU(2)_L$.
On top of these renormalizable terms, there are 15 dimension-six bosonic operators, which are presented in Table~\ref{tab:SMEFT_ops}. We use the convention $\widetilde{F}^{\mu \nu} = \frac{1}{2} \varepsilon^{\mu \nu \rho \sigma} F_{\rho \sigma}$ with $\varepsilon^{0123} = -\varepsilon_{0123} = +1$.

\begin{table}
    \centering
    \bgroup
    \def\arraystretch{1.5} 
    \begin{tabular}{|c|c|}
        \hline
         & $X^2 H^2$ \\
        \hline
        $C_{HB}$ & $H^\dag H \, B^{\mu \nu} \, B_{\mu \nu}$ \\
        $C_{HW}$ & $H^\dag H \, W^{I \, \mu \nu} \, W_{\mu \nu}^I$ \\
        $C_{HG}$ & $H^\dag H \, G^{A \, \mu \nu} \, G_{\mu \nu}^A$ \\
        $C_{H\widetilde{B}}$ & $H^\dag H \, B^{\mu \nu} \, \widetilde{B}_{\mu \nu}$ \\
        $C_{H\widetilde{W}}$ & $H^\dag H \, W^{I \, \mu \nu} \, \widetilde{W}_{\mu \nu}^I$ \\
        $C_{H\widetilde{G}}$ & $H^\dag H \, G^{A \, \mu \nu} \, \widetilde{G}_{\mu \nu}^A$ \\
        $C_{HWB}$ & $ \big( H^\dag \sigma^I H \big) B^{\mu \nu} \, W_{\mu \nu}^I$ \\
        $C_{H\widetilde{W}B}$ & $ \big(H^\dag \sigma^I H \big) B^{\mu \nu} \, \widetilde{W}_{\mu \nu}^I$ \\
        \hline
    \end{tabular}
    \quad
    \begin{tabular}{|c|c|}
        \hline
         & $X^3$ \\
        \hline
        $C_{W}$ & $f^{I J K} \, W_\mu^{I \, \nu} \, W_{\nu}^{J \, \rho} \, W^{K \, \mu}_\rho$ \\
        $C_{G}$ & $f^{A B C} \, G_\mu^{A \, \nu} \, G_{\nu}^{B \, \rho} \, G^{C \, \mu}_\rho$ \\
        $C_{\widetilde{W}}$ & $f^{I J K} \, \widetilde{W}_\mu^{I \, \nu} \, W_{\nu}^{J \, \rho} \, W^{K \, \mu}_\rho$ \\
        $C_{\widetilde{G}}$ & $f^{A B C} \, \widetilde{G}_\mu^{A \, \nu} \, G_{\nu}^{B \, \rho} \, G^{C \, \mu}_\rho$ \\
        \hline
         & $H^4 D^2$ and $H^6$ \\
         \hline
        $C_{H}$ & $\lzm H^\dag H \dzm^3$ \\
        $C_{H\Box}$ & $ \big(H^\dag H\big)  \Box \lzm H^\dag H \dzm$ \\
        $C_{HD}$ & $\lzm H^\dag D_\mu H \dzm^\ast \lzm H^\dag D^\mu H \dzm$ \\
        \hline
    \end{tabular}
    \egroup
    \caption{Dimension six bosonic SMEFT operators}
    \label{tab:SMEFT_ops}
\end{table}

As mentioned earlier, the calculations are performed in the background-field gauge so as to maintain manifest background gauge covariance at all intermediate steps. Each gauge field is split into a quantum field $A_\mu^a$ and a background field $\Hat{A}_\mu^a$ and the gauge-fixing Lagrangian is written as  
\begin{align}
    \L_\text{gf.} = -\frac{1}{2\xi} \lzm \partial^\mu A_\mu^a + g' f\ud{a}{bc} \Hat{A}^{b \, \mu} A^c_\mu \dzm^2 \equiv -\frac{1}{2\xi} \lzm \widehat{D}^\mu A_\mu^a \dzm^2 \, ,
\end{align}
where $\widehat{D}$ denotes the covariant derivative containing a background field $\Hat{A}_\mu^a$ and $\xi$ is the gauge-fixing parameter. Following the usual approach in functional calculations (see e.g.~\cite{Henning:2014wua}), we fix $ \xi = 1 $ for the Feynman gauge. The corresponding ghost Lagrangian is
\begin{align}
    \L_\text{gh.} = - \Bar{c}^a \widehat{D}^\mu \big(\widehat{D}_\mu c^a + f\ud{a}{bc} A^b_\mu c^c \big)\, ,
\end{align}
with the ghost fields $c^a$. 
Last but not least, the $\theta$ terms are given by
\begin{align}
    \L_\theta = \frac{\theta_1 g_Y^2}{32 \pi^2} \widetilde{B}_{\mu \nu} B^{\mu \nu} + \frac{\theta_2 g_L^2}{32 \pi^2} \widetilde{W}_{\mu \nu}^I W^{I \, \mu \nu} + \frac{\theta_3 g_s^2}{32 \pi^2} \widetilde{G}_{\mu \nu}^A G^{A \, \mu \nu} \, .
\end{align}

\subsection{Computer implementation}

Starting from the two-loop counterterm master formula~\eqref{eq:TwoLoopCT}, the computer implementation is in principle straightforward. However, just blindly hard-coding these formulae is highly inefficient as it leads to an unnecessarily large number of terms, many of which are redundant in the sense that they either evaluate to zero under integration, exceed the desired EFT order, or can be related to others using the symmetries discussed in Section~\ref{subsec:combinatorics}. The challenge, thus, lies in organizing the intermediate steps of the calculation in an efficient way that avoids this proliferation of terms. We found the following structuring to be efficient:
\begin{enumerate}
    \item \label{item:n-point} Derive the two-point ($Q_{ab}$), three-point ($C_{abc}$) and four-point ($D_{abcd}$) functions from the Lagrangian by applying functional derivatives.
    \item Determine all possible sunset and figure-8 types up to the desired EFT order, taking into account the symmetries described in Section~\ref{subsec:combinatorics}. At this stage, any field labels on the vertices and the dressed propagators are kept general as \textit{scalar} and \textit{vector}. For instance, we write the interaction term at the level of  $X_{SS}$ instead of already inserting all possible combinations of labels, e.g. $X_{H H}$, $X_{H H^\dag}$, $X_{H^\dag H}$ and $X_{H^\dag H^\dag}$.
    \item Replace the generic labels \textit{scalar} and \textit{vector} with the field labels, taking into account the symmetries discussed in Section~\ref{subsec:combinatorics}, and insert the explicit form of the corresponding quantities as determined in the first step. This means that we change from a generic description of the sunset and the figure-8 topologies in terms of the proxy variables $X_{ab}$, $C_{abc}$, and $D_{abcd}$ to their model-specific form calculated in step~\ref{item:n-point}.
    \item Discard all integrals that are not logarithmically divergent. They have vanishing counterterms, since all integrals are scaleless after expansion of $ \mathcal{Q}^{\eminus 1} $ in the large momentum region.
    \item Move open derivatives from the vertex (or vertices) to the propagators.
    \item Compute the resulting integrals with the $\boldsymbol{R}^{\ast} $-method and dummy masses for IR regulation as described in Section~\ref{sec:RStar}.
    \item Evaluate derivatives on Wilson lines through repeated use of Eq.~\eqref{eq:D_on_U_right}.
\end{enumerate}

Working in the functional formalism, the output of this procedure is an unsimplified counterterm Lagrangian. This can be brought to an on-shell (or off-shell) EFT basis using the simplification functionality of \texttt{Matchete}~\cite{Fuentes-Martin:2022jrf}, thereby recovering the Warsaw-basis counterterms.

\subsection{Two-loop results}
Given the highly-complex and automated calculation, validation of the implementation and results are crucial to establishing their soundness. Our code---built on top of the model-generic \texttt{Matchete} codebase---can easily be applied to models other than the bosonic SMEFT.\footnote{Although our implementation is presently exclusive to bosonic theories.} This allows us to validate our implementation against other models, which is a great help considering how few two-loop results are available for bosonic EFTs. Thus, we have validated our implementation against the following known results:
\begin{enumerate}[i)]
    \item the one-loop \befs of the SMEFT~\cite{Jenkins:2013zja,Alonso:2013hga};
    \item the two-loop counterterms of the scalar toy model of~\cite{Fuentes-Martin:2023ljp}, which was calculated by hand;
    \item the two-loop \befs and field anomalous dimension in renormalizable scalar QED and scalar QCD as calculated with \texttt{RGBeta}~\cite{Thomsen:2021ncy}; 
    \item the two-loop counterterms of the scalar $O(N)$ model (dimension six)~\cite{Jenkins:2023bls} with $N=5$ and $N=6$;
    \item the two-loop counterterms of the SMEFT scalar sector~\cite{Jenkins:2023bls};
    \item the two-loop counterterms in a Yang-Mills theory with a Weinberg operator~\cite{deVries:2019nsu}.
\end{enumerate}
The Weinberg operator, in particular, provides a very stringent check given the complexity of the $ G G \widetilde{G}$-vertex; this is as complex as the bosonic SMEFT gets. Additionally, this calculation involves highly-complicated numerators in the Feynman integrals with many momentum powers and Lorentz index contractions, supporting our implementation of the $\boldsymbol{R}^{\ast} $-method.

Furthermore, we confirmed that the double poles of the counterterms we calculated for the bosonic SMEFT fulfill the `t Hooft consistency condition
\begin{align} \label{eq:ConsistencyCheck}
    \zeta \delta g_{I(n+1)} = \beta_J \partial^J \delta g_{In} \, ,\qquad \partial^I \equiv \dfrac{\partial}{\partial g_I},
\end{align}
where $\zeta$ is the loop-counting operator ($\zeta(\hbar^n) = 2n \hbar^n $), $g_{In}$ denotes the $n$'th pole part in the counterterm associated to the coupling $g_I$, and $\beta_J$ is the \bef of $g_J$. Hence, the two-loop double poles are determined uniquely by the one-loop single poles. We find full agreement between our explicit counterterm calculations and the consistency condition for all couplings in the bosonic SMEFT. Even though fulfilling the condition does not ensure that the simple poles and, therefore, the two-loop \befs are correct, it still provides a strong crosscheck of our results. From the single pole of the two-loop counterterms, we have computed the \befs for all couplings in the bosonic SMEFT. These beta functions are given in Appendix~\ref{app:betas} and in the ancillary file \texttt{betaFunctions.m}.

\section{Conclusion and Outlook} \label{sec:conclusion}
Two-loop SMEFT running opens up several important phenomenological effects in both top--down studies of BSM models and bottom--up global analyses. Additionally, the inclusion of the two-loop running should reduce the unphysical scale dependence in theory predictions. The SMEFT is a complicated model already at dimension six and it is a serious challenge to calculate the \befs of all its Wilson coefficients. The functional framework has proven very useful for organizing large calculations in, for instance, one-loop EFT matching. Thus, it stands to reason that the development of functional methods for multi-loop counterterm calculations can benefit the calculation of two-loop SMEFT \befs. 

The main new developments of the methods made in the present paper were the adaptation of the $\boldsymbol{R^\ast}$-method to the functional formulas; it justifies the expansion of the dressed propagators while conveniently subtracting any subdivergences. In this approach, there is no need to explicitly determine all one-loop counterterms off-shell to manually subtract UV subdivergences, nor are there any issues with IR subdivergences. The implementation of functional multi-loop methods in a computer code also serves to demonstrate their practicality for application to phenomenologically relevant models. 
Using our implementation, we present for the first time the complete two-loop \befs for the bosonic SMEFT. While this is but a partial result as far as phenomenology is concerned, it represents an important milestone towards the full evaluation. 

As demonstrated implicitly by our code validation, our implementation is already generic to the extent that the specific model can be changed effortlessly. 
It is our intention to build on this work by including fermions, thereby massively extending the space of models covered. In this event, it would be possible to obtain the full two-loop running results for the entire SMEFT. 
With further refinement, it could also become possible to distribute the two-loop capabilities with the \texttt{Matchete} package.

\subsection*{Acknowledgments}
AET would like to thank Luca Naterop and Peter Stoffer for many helpful discussions. 
The work of LB and AET is funded by the Swiss National Science Foundation (SNSF) through the Ambizione grant ``Matching and Running: Improved Precision in the Hunt for New Physics,'' project number 209042. The work of JFM is supported by the grants IJC2020-043549-I and EUR2024.153549 funded by MCIN/AEI/10.13039/ 501100011033 and the European Union NextGenerationEU/PRTR, PID2022-139466NB-C21 funded by MICIU/AEI/10.13039/501100011033 and FEDER/UE, and by the Junta de Andaluc\'ia grants P21\_00199 and FQM101. 
The work of SK has been funded by the
European Union NextGenerationEU/PRTR under grant CNS2022-136024 and by MICIU/AEI/10.13039/501100011033 and FEDER/UE
(grant PID2022-139466NB-C21). The work of SK is further supported by the Junta de Andaluc\'ia grant AST22\_6.5.

\renewcommand{\thesection}{\Alph{section}}
\appendix 

\section{Two-loop \texorpdfstring{\befs}{beta functions} of the bosonic SMEFT} \label{app:betas}

In this appendix, we list the beta-functions of the bosonic SMEFT obtained with functional methods. For a coupling $X$, we parameterize the beta-functions by
\begin{equation}
    \beta_X = \frac{\text{d}X}{\text{d}\ln\bar{\mu}}=\sum_{\ell=1}^\infty \frac{\beta_X^{(\ell)}}{(4\pi)^{2\ell}},
\end{equation}
where $\bar{\mu}$ is the $\overline{\text{MS}}$ renormalization scale.

The \befs of the renormalizable couplings in the presence of dimension-six operators are given by
\begin{align}
\begin{split} 
\beta_{ \mu ^{ 2} }^{(1)}=\,& \mu ^{ 2} \Big( 6 \lambda -  \frac{3}{2} g_Y ^{ 2} -  \frac{9}{2} g_L ^{ 2} \Big) +  \mu ^{ 4} \Big( 4 C_{HD} -  8 C_{H\Box} \Big), \end{split} 
\\[5pt] 
\begin{split} 
\beta_{ \mu ^{ 2} }^{(2)}=\,& \mu ^{ 2} \Big[ \lambda \Big( 12 g_Y ^{ 2} +  36 g_L ^{ 2} \Big) +  \frac{15}{8} g_Y ^{ 2} g_L ^{ 2} +  \frac{157}{48} g_Y ^{ 4} -  15 \lambda ^{ 2} -  \frac{385}{16} g_L ^{ 4} \Big]  \\[2pt] 
 &+  \mu ^{ 4} \Big[ \Big( 6 g_Y ^{ 2} +  24 g_L ^{ 2} -  20 \lambda \Big) C_{HD} +  \Big( 80 \lambda -  18 g_Y ^{ 2} -  54 g_L ^{ 2} \Big) C_{H\Box}  \\ 
 &+  15 g_L ^{ 3} C_W +  18 g_Y g_L C_{HWB} -  10 g_Y ^{ 2} C_{HB} -  30 g_L ^{ 2} C_{HW} \Big], \end{split}
\\
\nonumber
\\
\begin{split} 
\beta_{ \lambda }^{(1)}=\,&\frac{3}{4} g_Y ^{ 4} +  \lambda \Big( - 3 g_Y ^{ 2} -  9 g_L ^{ 2} \Big) +  \mu ^{ 2} \Big[ \Big( \frac{40}{3} g_L ^{ 2} -  64 \lambda \Big) C_{H\Box} +  \Big( 6 g_Y ^{ 2} +  24 \lambda -  6 g_L ^{ 2} \Big) C_{HD}  \\ 
 & +  12 g_Y g_L C_{HWB} +  12 g_Y ^{ 2} C_{HB} +  36 g_L ^{ 2} C_{HW} +  48 C_H \Big]  +  \frac{3}{2} g_Y ^{ 2} g_L ^{ 2} +  \frac{9}{4} g_L ^{ 4} +  12 \lambda ^{ 2} , 
\end{split} 
\\[5pt]  
\begin{split} 
\beta_{ \lambda }^{(2)}=\,& \lambda \Big( \frac{229}{24} g_Y ^{ 4} +  \frac{39}{4} g_Y ^{ 2} g_L ^{ 2} -  \frac{313}{8} g_L ^{ 4} \Big) +  \lambda ^{ 2} \Big( 18 g_Y ^{ 2} +  54 g_L ^{ 2} \Big)  \\ 
 &+  \mu ^{ 2} \Big\{ \Big( 96 g_Y ^{ 2} +  288 g_L ^{ 2} -  480 \lambda \Big) C_H +  \Big( \frac{1121}{3} g_L ^{ 4} -  65 g_Y ^{ 2} g_L ^{ 2} -  348 \lambda g_L ^{ 2} \Big) C_{HW} \\ 
 & +  \Big( 21 g_Y ^{ 2} g_L ^{ 3} +  108 \lambda g_L ^{ 3} -  \frac{413}{3} g_L ^{ 5} \Big) C_W +  \Big( 16 g_Y g_L ^{ 3} +  92 \lambda g_Y g_L -  \frac{434}{3} g_Y ^{ 3} g_L \Big) C_{HWB}  \\ 
 & +  \Big( - \frac{163}{3} g_Y ^{ 2} g_L ^{ 2} -  84 \lambda g_Y ^{ 2} -  \frac{283}{3} g_Y ^{ 4} \Big) C_{HB} +  \Big[ \lambda \Big( 9 g_Y ^{ 2} +  159 g_L ^{ 2} \Big) -  \frac{239}{24} g_L ^{ 4} \\ 
 & -  \frac{161}{8} g_Y ^{ 4}  -  \frac{5209}{108} g_Y ^{ 2} g_L ^{ 2} -  282 \lambda ^{ 2} \Big] C_{HD}  +  \Big[ \lambda \Big( - 132 g_Y ^{ 2} -  \frac{1460}{3} g_L ^{ 2} \Big) +  \frac{347}{6} g_L ^{ 4}   \\ 
 &+  1096 \lambda ^{ 2} -  \frac{859}{27} g_Y ^{ 2} g_L ^{ 2} -  \frac{217}{6} g_Y ^{ 4} \Big] C_{H\Box} \Big\} +  \frac{497}{8} g_L ^{ 6} -  \frac{59}{24} g_Y ^{ 6} -  \frac{97}{24} g_Y ^{ 2} g_L ^{ 4}  \\ 
 &-  \frac{239}{24} g_Y ^{ 4} g_L ^{ 2} -  78 \lambda ^{ 3} , 
\end{split}
\\
\nonumber
\\
\begin{split} 
\beta_{ g_Y ^{ 2} }^{(1)}=\,&\frac{1}{3} g_Y ^{ 4} -  16 \mu ^{ 2} g_Y ^{ 2} C_{HB}, 
\end{split} 
 \\[5pt] 
\begin{split} 
\beta_{ g_Y ^{ 2} }^{(2)}=\,& g_Y ^{ 6} +  \mu ^{ 2} \Big[ \Big( - 24 g_Y ^{ 4} -  96 g_Y ^{ 2} g_L ^{ 2} \Big) C_{HB} +  12 g_Y ^{ 3} g_L C_{HWB} \Big] +  3 g_Y ^{ 4} g_L ^{ 2} , 
\end{split}
\\
\nonumber
\\
\begin{split} 
\beta_{ g_L ^{ 2} }^{(1)}=\,&- \frac{43}{3} g_L ^{ 4} -  16 \mu ^{ 2} g_L ^{ 2} C_{HW}, 
\end{split} 
 \\ 
\begin{split} 
\beta_{ g_L ^{ 2} }^{(2)}=\,& g_Y ^{ 2} g_L ^{ 4} +  \mu ^{ 2} \Big[ \Big( - 32 g_Y ^{ 2} g_L ^{ 2} -  88 g_L ^{ 4} \Big) C_{HW} +  4 g_Y g_L ^{ 3} C_{HWB} -  228 g_L ^{ 5} C_W \Big]  -  \frac{259}{3} g_L ^{ 6} , 
\end{split}
\\
\nonumber
\\
\begin{split} 
\beta_{ g_s ^{ 2} }^{(1)}=\,&- 16 \mu ^{ 2} g_s ^{ 2} C_{HG} -  22 g_s ^{ 4} , 
\end{split} 
\\[5pt] 
\begin{split} 
\beta_{ g_s ^{ 2} }^{(2)}=\,& \mu ^{ 2} g_s ^{ 2} \Big( - 32 g_Y ^{ 2} -  96 g_L ^{ 2} \Big) C_{HG} -  204 g_s ^{ 6} , 
\end{split}
\\
\nonumber
\\
\begin{split}
\beta_{\theta_1}^{(0)} =\,& - 8 \frac{\mu ^2}{g_Y^2} C_{H \widetilde{B}} 
,
\end{split}
\\[5pt]
\begin{split}
\beta_{\theta_1}^{(1)} =\,& 6 \frac{\mu ^2 g_{L}}{g_Y} C_{H \widetilde{W} B} + \mu^2 \Big( - 48  \frac{g_{L}^2}{g_Y^2} - 12 \Big) C_{H \widetilde{B}}
,
\end{split}
\\
\nonumber
\\
\begin{split}
\beta_{\theta_2}^{(0)} =\,& -8 \frac{\mu^2}{g_L^2}  C_{H \widetilde{W}}
,
\end{split}
\\[5pt]
\begin{split}
\beta_{\theta_2}^{(1)} =\,& 2\mu ^2 \frac{g_Y}{g_L} C_{H \widetilde{W} B} + \mu^2 \Big(-16\frac{g_Y^2}{g_L^2}-44 \Big) C_{H \widetilde{W}} - 74 \mu^2 g_L C_{\widetilde{W}}
,
\end{split}
\\
\nonumber
\\
\begin{split}
\beta_{\theta_3}^{(0)} =\,& -8\frac{\mu^2}{g_s^2} C_{H \tilde{G}}
,
\end{split}
\\[5pt]
\begin{split}
\beta_{\theta_3}^{(1)} =\,& \mu^2 \Big( - 48 \frac{g_L^2}{g_s^2} - 16 \frac{g_Y^2}{g_s^2}\Big) C_{H \tilde{G}} \, .
\end{split}
\end{align}
Including a loop factor $\frac{1}{16\pi^2}$ in the definition of the $\theta$ term formally reduces the loop order of the corresponding \befs by one, therefore we write $\beta_{\theta_i}^{(0)}$ and $\beta_{\theta_i}^{(1)}$ for the one-loop and two-loop \befs, respectively. 

The \befs for the Wilson coefficients of the dimension-six operators are 
\begin{align}
\begin{split} 
\beta_{ C_H}^{(1)}=\,& \Big( 54 \lambda -  \frac{9}{2} g_Y ^{ 2} -  \frac{27}{2} g_L ^{ 2} \Big) C_H +  \Big( 18 \lambda g_L ^{ 2} -  3 g_Y ^{ 2} g_L ^{ 2} -  9 g_L ^{ 4} \Big) C_{HW}  \\ 
 &+  \Big( 6 \lambda g_Y g_L -  3 g_Y g_L ^{ 3} -  3 g_Y ^{ 3} g_L \Big) C_{HWB} +  \Big( 6 \lambda g_Y ^{ 2} -  3 g_Y ^{ 2} g_L ^{ 2} -  3 g_Y ^{ 4} \Big) C_{HB}  \\ 
 &+  \Big( \frac{20}{3} \lambda g_L ^{ 2} -  40 \lambda ^{ 2} \Big) C_{H\Box} +  \Big[ \lambda \Big( 3 g_Y ^{ 2} -  3 g_L ^{ 2} \Big) +  12 \lambda ^{ 2} -  \frac{3}{4} g_L ^{ 4}  \\ 
 &-  \frac{3}{4} g_Y ^{ 4} -  \frac{3}{2} g_Y ^{ 2} g_L ^{ 2} \Big] C_{HD}, 
\end{split} 
\\[5pt] 
\begin{split} 
\beta_{ C_H}^{(2)}=\,& \Big[ \lambda \Big( 72 g_Y ^{ 2} +  216 g_L ^{ 2} \Big) +  \frac{301}{16} g_Y ^{ 4} +  \frac{189}{8} g_Y ^{ 2} g_L ^{ 2} -  \frac{723}{16} g_L ^{ 4} -  861 \lambda ^{ 2} \Big] C_H\\
 &+  \Big[ \lambda \Big( \frac{617}{6} g_L ^{ 4} -  \frac{121}{2} g_Y ^{ 2} g_L ^{ 2} \Big) +  \frac{335}{12} g_Y ^{ 4} g_L ^{ 2} +  \frac{193}{6} g_Y ^{ 2} g_L ^{ 4} -  180 \lambda ^{ 2} g_L ^{ 2}\\
 &-  \frac{1395}{4} g_L ^{ 6} \Big] C_{HW} +  \Big[\lambda \Big( \frac{21}{2} g_Y ^{ 2} g_L ^{ 3} -  \frac{413}{6} g_L ^{ 5} \Big) +  6 g_Y ^{ 2} g_L ^{ 5} +  39 \lambda ^{ 2} g_L ^{ 3}\\
 &+  63 g_L ^{ 7} -  6 g_Y ^{ 4} g_L ^{ 3} \Big] C_W +  \Big[ \Big( - 20 g_Y g_L ^{ 3} -  \frac{301}{3} g_Y ^{ 3} g_L \Big) \lambda +  16 \lambda ^{ 2} g_Y g_L\\
 &+  \frac{319}{12} g_Y ^{ 5} g_L +  \frac{111}{2} g_Y ^{ 3} g_L ^{ 3} -  \frac{409}{12} g_Y g_L ^{ 5} \Big] C_{HWB} +  \Big[ \lambda \Big( - \frac{331}{6} g_Y ^{ 2} g_L ^{ 2}  \\ 
 &-  \frac{451}{6} g_Y ^{ 4} \Big) +  \frac{193}{12} g_Y ^{ 2} g_L ^{ 4} +  \frac{91}{4} g_Y ^{ 6} +  \frac{335}{6} g_Y ^{ 4} g_L ^{ 2} -  44 \lambda ^{ 2} g_Y ^{ 2} \Big] C_{HB}  \\ 
 &+  \Big[ \lambda \Big( - \frac{95}{48} g_L ^{ 4} -  \frac{209}{16} g_Y ^{ 4} -  \frac{6505}{216} g_Y ^{ 2} g_L ^{ 2} \Big) +  \lambda ^{ 2} \Big( \frac{123}{2} g_L ^{ 2} -  \frac{3}{2} g_Y ^{ 2} \Big)  \\ 
 &+  \frac{59}{12} g_Y ^{ 6} +  14 g_Y ^{ 2} g_L ^{ 4} +  \frac{149}{8} g_Y ^{ 4} g_L ^{ 2} -  \frac{497}{24} g_L ^{ 6} -  249 \lambda ^{ 3} \Big] C_{HD}  \\ 
 &+  \Big[ \lambda \Big( - \frac{85}{12} g_L ^{ 4} -  \frac{361}{12} g_Y ^{ 4} -  \frac{2155}{54} g_Y ^{ 2} g_L ^{ 2} \Big) +  \lambda ^{ 2} \Big( - 60 g_Y ^{ 2} -  \frac{676}{3} g_L ^{ 2} \Big)  \\ 
 &+  \frac{13}{3} g_Y ^{ 6} +  13 g_L ^{ 6} +  13 g_Y ^{ 2} g_L ^{ 4} +  13 g_Y ^{ 4} g_L ^{ 2} +  996 \lambda ^{ 3} \Big] C_{H\Box},
\end{split}
\\
\nonumber
\\
\begin{split} \beta_{ C_{H\Box}}^{(1)}=\,& \Big( 12 \lambda -  \frac{4}{3} g_Y ^{ 2} -  4 g_L ^{ 2} \Big) C_{H\Box} +  \frac{5}{3} g_Y ^{ 2} C_{HD}, \end{split} 
 \\ 
\begin{split} 
\beta_{ C_{H\Box}}^{(2)}=\,& \Big( 19 g_Y ^{ 2} g_L ^{ 2} -  41 g_L ^{ 4} -  108 \lambda g_L ^{ 2} \Big) C_{HW} +  \Big( 2 g_Y ^{ 2} g_L ^{ 3} +  36 \lambda g_L ^{ 3}  \\ 
 &-  28 g_L ^{ 5} \Big) C_W +  \Big( 4 g_Y ^{ 3} g_L +  16 g_Y g_L ^{ 3} -  12 \lambda g_Y g_L \Big) C_{HWB} +  \Big( \frac{29}{3} g_Y ^{ 4}  \\ 
 &+  19 g_Y ^{ 2} g_L ^{ 2} -  36 \lambda g_Y ^{ 2} \Big) C_{HB} +  \Big( - \frac{23}{27} g_Y ^{ 4} -  \frac{65}{36} g_Y ^{ 2} g_L ^{ 2} -  \frac{20}{3} \lambda g_Y ^{ 2} \Big) C_{HD}  \\ 
 &+  \Big[ \lambda \Big( 11 g_Y ^{ 2} +  33 g_L ^{ 2} \Big) +  \frac{1945}{432} g_Y ^{ 4} +  \frac{99}{16} g_L ^{ 4} +  \frac{523}{72} g_Y ^{ 2} g_L ^{ 2}  -  51 \lambda ^{ 2} \Big] C_{H\Box}, 
\end{split}
\\
\nonumber
\\
\begin{split} 
\beta_{ C_{HD}}^{(1)}=\,& \Big( \frac{9}{2} g_L ^{ 2} +  6 \lambda -  \frac{5}{6} g_Y ^{ 2} \Big) C_{HD} +  \frac{20}{3} g_Y ^{ 2} C_{H\Box}, 
\end{split} 
\\[5pt]
\begin{split} 
\beta_{ C_{HD}}^{(2)}=\,& \Big( 32 g_Y ^{ 3} g_L -  68 g_Y g_L ^{ 3} -  96 \lambda g_Y g_L \Big) C_{HWB} +  \Big( \frac{32}{3} g_Y ^{ 4} +  12 g_Y ^{ 2} g_L ^{ 2} -  48 \lambda g_Y ^{ 2} \Big) C_{HB}  \\ 
 &+  \Big( \frac{70}{27} g_Y ^{ 4} -  \frac{227}{9} g_Y ^{ 2} g_L ^{ 2} -  \frac{136}{3} \lambda g_Y ^{ 2} \Big) C_{H\Box} +  \Big[\lambda \Big( \frac{5}{2} g_Y ^{ 2} -  \frac{45}{2} g_L ^{ 2} \Big) +  \frac{299}{216} g_Y ^{ 4}  \\ 
 &+  \frac{41}{2} g_Y ^{ 2} g_L ^{ 2} -  \frac{1}{8} g_L ^{ 4} -  36 \lambda ^{ 2} \Big] C_{HD} +  26 g_Y ^{ 2} g_L ^{ 3} C_W +  28 g_Y ^{ 2} g_L ^{ 2} C_{HW}, 
 \end{split}
\\
\nonumber
\\
\begin{split} 
\beta_{ C_{HB}}^{(1)}=\,& \Big( \frac{5}{6} g_Y ^{ 2} +  6 \lambda -  \frac{9}{2} g_L ^{ 2} \Big) C_{HB} +  3 g_Y g_L C_{HWB}, 
\end{split} 
 \\[5pt] 
\begin{split} 
\beta_{ C_{HB}}^{(2)}=\,& \Big( \frac{3}{2} g_Y ^{ 3} g_L +  \frac{125}{6} g_Y g_L ^{ 3} -  9 \lambda g_Y g_L \Big) C_{HWB} +  \Big( \frac{8}{3} g_Y ^{ 2} g_L ^{ 2} -  \frac{1}{9} g_Y ^{ 4} -  2 \lambda g_Y ^{ 2} \Big) C_{H\Box}  \\ 
 &+  \Big( \frac{23}{36} g_Y ^{ 4} -  \frac{3}{4} g_Y ^{ 2} g_L ^{ 2} -  \lambda g_Y ^{ 2} \Big) C_{HD} +  \Big( \frac{739}{144} g_Y ^{ 4} +  \frac{75}{8} g_Y ^{ 2} g_L ^{ 2} +  36 \lambda g_L ^{ 2}  \\ 
 &-  15 \lambda ^{ 2} -  \frac{385}{16} g_L ^{ 4} \Big) C_{HB} +  \frac{9}{4} g_Y ^{ 2} g_L ^{ 2} C_{HW} -  \frac{45}{8} g_Y ^{ 2} g_L ^{ 3} C_W, 
\end{split}
\\
\nonumber
\\
\begin{split} 
\beta_{ C_{HG}}^{(1)}=\,& \Big( 6 \lambda -  \frac{3}{2} g_Y ^{ 2} -  \frac{9}{2} g_L ^{ 2} -  22 g_s ^{ 2}  \Big)  C_{HG}, 
\end{split} 
\\[5pt] 
\begin{split} 
\beta_{ C_{HG}}^{(2)}=\,& \Big[  \lambda \Big( 12 g_Y ^{ 2} +  36 g_L ^{ 2} \Big) +  \frac{15}{8} g_Y ^{ 2} g_L ^{ 2} +  \frac{157}{48} g_Y ^{ 4} -  15 \lambda ^{ 2} -  \frac{385}{16} g_L ^{ 4} -  408 g_s ^{ 4}  \Big]  C_{HG}, 
\end{split}
\\
\nonumber
\\
\begin{split} 
\beta_{ C_G}^{(1)}=\,&3  g_s ^{ 2}   C_G, 
\end{split} 
\\[4pt] 
\begin{split} 
\beta_{ C_G}^{(2)}=\,&75  g_s ^{ 4}   C_G, 
\end{split}
\\
\nonumber
\\
\begin{split} 
\beta_{ C_{HWB}}^{(1)}=\,& \Big( 2 \lambda -  \frac{1}{3} g_Y ^{ 2} -  \frac{8}{3} g_L ^{ 2} \Big) C_{HWB} +  2 g_Y g_L C_{HB} +  2 g_Y g_L C_{HW} -  3 g_Y g_L ^{ 2} C_W, 
\end{split} 
\\[4pt]  
\begin{split} 
\beta_{ C_{HWB}}^{(2)}=\,& \Big( \frac{1}{2} g_Y ^{ 3} g_L -  \frac{11}{18} g_Y g_L ^{ 3} -  6 \lambda g_Y g_L \Big) C_{HW} +  \Big( \frac{11}{18} g_Y ^{ 3} g_L -  \frac{11}{2} g_Y g_L ^{ 3}  \\ 
 &-  6 \lambda g_Y g_L \Big) C_{HB} +  \Big( \frac{7}{9} g_Y ^{ 3} g_L +  g_Y g_L ^{ 3} -  4 \lambda g_Y g_L \Big) C_{H\Box} +  \Big( \frac{1}{36} g_Y ^{ 3} g_L  \\ 
 &+  \frac{1}{4} g_Y g_L ^{ 3} -  \lambda g_Y g_L \Big) C_{HD} +  \Big( - \frac{3}{8} g_Y ^{ 3} g_L ^{ 2} -  \frac{9}{2} \lambda g_Y g_L ^{ 2} -  \frac{169}{24} g_Y g_L ^{ 4} \Big) C_W  \\ 
 &+  \Big( \frac{13}{48} g_L ^{ 4} +  \frac{263}{144} g_Y ^{ 4} +  \frac{91}{8} g_Y ^{ 2} g_L ^{ 2} -  2 \lambda g_L ^{ 2} -  7 \lambda ^{ 2} \Big) C_{HWB},
\end{split}
\\
\nonumber
\\
\begin{split} 
\beta_{ C_{HW}}^{(1)}=\,& g_Y g_L C_{HWB} +  \Big( 6 \lambda -  \frac{3}{2} g_Y ^{ 2} -  \frac{101}{6} g_L ^{ 2} \Big) C_{HW} +  15 g_L ^{ 3} C_W, \end{split} 
\\[5pt]  
\begin{split} 
\beta_{ C_{HW}}^{(2)}=\,&\frac{7}{18} g_Y ^{ 2} g_L ^{ 2} C_{HD} +  \frac{3}{4} g_Y ^{ 2} g_L ^{ 2} C_{HB} +  \Big( \frac{8}{9} g_Y ^{ 2} g_L ^{ 2} +  g_L ^{ 4} -  2 \lambda g_L ^{ 2} \Big) C_{H\Box}  \\ 
 &+  \Big( \frac{171}{2} \lambda g_L ^{ 3} +  \frac{1859}{6} g_L ^{ 5} -  \frac{177}{8} g_Y ^{ 2} g_L ^{ 3} \Big) C_W +  \Big( \frac{1}{18} g_Y ^{ 3} g_L +  10 g_Y g_L ^{ 3}  \\ 
 &-  3 \lambda g_Y g_L \Big) C_{HWB} +  \Big[ \lambda \Big( 12 g_Y ^{ 2} +  24 g_L ^{ 2} \Big) +  \frac{157}{48} g_Y ^{ 4} +  \frac{35}{8} g_Y ^{ 2} g_L ^{ 2}  \\ 
 &-  15 \lambda ^{ 2} -  \frac{23701}{144} g_L ^{ 4} \Big] C_{HW}, 
 \end{split}
\\
\nonumber
\\
\begin{split} 
\beta_{ C_W}^{(1)}=\,&\frac{5}{2}  g_L ^{ 2}   C_W, 
\end{split} 
\\[5pt]  
\begin{split} 
\beta_{ C_W}^{(2)}=\,& \Big( \frac{3}{2} g_Y ^{ 2} g_L ^{ 2} +  \frac{181}{6} g_L ^{ 4} \Big) C_W +  5 g_L ^{ 3} C_{HW} -  \frac{1}{2} g_Y g_L ^{ 2} C_{HWB}, 
\end{split}
\\
\nonumber
\\
\begin{split} 
\beta_{ C_{H\widetilde{B}}}^{(1)}=\,& \Big( \frac{5}{6} g_Y ^{ 2} +  6 \lambda -  \frac{9}{2} g_L ^{ 2} \Big) C_{H\widetilde{B}} +  3 g_Y g_L C_{H\widetilde{W}B}, 
\end{split} 
\\[5pt] 
\begin{split} \beta_{ C_{H\widetilde{B}}}^{(2)}=\,& \Big( \frac{3}{2} g_Y ^{ 3} g_L +  \frac{125}{6} g_Y g_L ^{ 3} -  9 \lambda g_Y g_L \Big) C_{H\widetilde{W}B} +  \Big( \frac{595}{144} g_Y ^{ 4} +  \frac{75}{8} g_Y ^{ 2} g_L ^{ 2} +  36 \lambda g_L ^{ 2}  \\ 
 &-  15 \lambda ^{ 2} -  \frac{385}{16} g_L ^{ 4} \Big) C_{H\widetilde{B}} -  \frac{3}{4} g_Y ^{ 2} g_L ^{ 2} C_{H\widetilde{W}} -  \frac{21}{8} g_Y ^{ 2} g_L ^{ 3} C_{\widetilde{W}}, \end{split}
\\
\nonumber
\\
\begin{split} \beta_{ C_{H\widetilde{G}}}^{(1)}=\,& \Big( 6 \lambda -  \frac{3}{2} g_Y ^{ 2} -  \frac{9}{2} g_L ^{ 2} -  22 g_s ^{ 2}  \Big)  C_{H\widetilde{G}}, \end{split} 
\\[5pt] 
\begin{split} 
\beta_{ C_{H\widetilde{G}}}^{(2)}=\,& \Big[  \lambda \Big( 12 g_Y ^{ 2} +  36 g_L ^{ 2} \Big) +  \frac{15}{8} g_Y ^{ 2} g_L ^{ 2} +  \frac{157}{48} g_Y ^{ 4} -  15 \lambda ^{ 2} -  \frac{385}{16} g_L ^{ 4} -  204 g_s ^{ 4}  \Big]  C_{H\widetilde{G}}, 
\end{split}
\\
\nonumber
\\
\begin{split} 
\beta_{ C_{H\widetilde{W}B}}^{(1)}=\,& \Big( 2 \lambda -  \frac{1}{3} g_Y ^{ 2} -  \frac{8}{3} g_L ^{ 2} \Big) C_{H\widetilde{W}B} +  2 g_Y g_L C_{H\widetilde{B}} +  2 g_Y g_L C_{H\widetilde{W}} -  3 g_Y g_L ^{ 2} C_{\widetilde{W}}, 
\end{split} 
\\[5pt] 
\begin{split} 
\beta_{ C_{H\widetilde{W}B}}^{(2)}=\,& \Big( \frac{1}{2} g_Y ^{ 3} g_L +  \frac{277}{18} g_Y g_L ^{ 3} -  6 \lambda g_Y g_L \Big) C_{H\widetilde{W}} +  \Big( \frac{11}{18} g_Y ^{ 3} g_L -  \frac{11}{2} g_Y g_L ^{ 3}  \\ 
 &-  6 \lambda g_Y g_L \Big) C_{H\widetilde{B}} +  \Big( - \frac{7}{8} g_Y ^{ 3} g_L ^{ 2} -  \frac{5}{2} \lambda g_Y g_L ^{ 2} -  \frac{95}{8} g_Y g_L ^{ 4} \Big) C_{\widetilde{W}}   \\ 
 &+  \Big( \frac{13}{48} g_L ^{ 4}+  \frac{263}{144} g_Y ^{ 4} +  \frac{75}{8} g_Y ^{ 2} g_L ^{ 2} -  2 \lambda g_L ^{ 2} -  7 \lambda ^{ 2} \Big) C_{H\widetilde{W}B}, 
\end{split}
\\
\nonumber
\\
\begin{split} 
\beta_{ C_{H\widetilde{W}}}^{(1)}=\,& g_Y g_L C_{H\widetilde{W}B} +  \Big( 6 \lambda -  \frac{3}{2} g_Y ^{ 2} -  \frac{101}{6} g_L ^{ 2} \Big) C_{H\widetilde{W}} +  15 g_L ^{ 3} C_{\widetilde{W}}, 
\end{split} 
\\ 
\begin{split} 
\beta_{ C_{H\widetilde{W}}}^{(2)}=\,& \Big( \frac{111}{2} \lambda g_L ^{ 3} +  199 g_L ^{ 5} -  \frac{109}{8} g_Y ^{ 2} g_L ^{ 3} \Big) C_{\widetilde{W}} +  \Big( \frac{1}{18} g_Y ^{ 3} g_L +  2 g_Y g_L ^{ 3} -  3 \lambda g_Y g_L \Big) C_{H\widetilde{W}B}  \\ 
 &+  \Big[ \lambda \Big( 12 g_Y ^{ 2} +  24 g_L ^{ 2} \Big) +  \frac{157}{48} g_Y ^{ 4} +  \frac{35}{8} g_Y ^{ 2} g_L ^{ 2} -  15 \lambda ^{ 2} -  \frac{11269}{144} g_L ^{ 4} \Big] C_{H\widetilde{W}}  \\ 
 &-  \frac{1}{4} g_Y ^{ 2} g_L ^{ 2} C_{H\widetilde{B}}, 
\end{split}
\\
\nonumber
\\
\begin{split} 
\beta_{ C_{\widetilde{G}}}^{(1)}=\,&3  g_s ^{ 2}   C_{\widetilde{G}}, 
\end{split} 
 \\[5pt] 
\begin{split} 
\beta_{ C_{\widetilde{G}}}^{(2)}=\,&119  g_s ^{ 4}   C_{\widetilde{G}}, 
\end{split}
\\
\nonumber
\\
\begin{split} 
\beta_{ C_{\widetilde{W}}}^{(1)}=\,&\frac{5}{2}  g_L ^{ 2}   C_{\widetilde{W}}, 
\end{split} 
 \\[5pt] 
\begin{split} 
\beta_{ C_{\widetilde{W}}}^{(2)}=\,& \Big( \frac{3}{2} g_Y ^{ 2} g_L ^{ 2} +  \frac{887}{18} g_L ^{ 4} \Big) C_{\widetilde{W}} +  5 g_L ^{ 3} C_{H\widetilde{W}} -  \frac{1}{2} g_Y g_L ^{ 2} C_{H\widetilde{W}B}. 
\end{split}
\end{align}

\sectionlike{References}
\vspace{-10pt}
\bibliography{References}

\providecommand{\href}[2]{#2}\begingroup\raggedright\begin{thebibliography}{10}

\bibitem{Feruglio:2016gvd}
F.~Feruglio, P.~Paradisi and A.~Pattori, \emph{{Revisiting Lepton Flavor
  Universality in B Decays}},
  \href{https://doi.org/10.1103/PhysRevLett.118.011801}{\emph{Phys. Rev. Lett.}
  {\bfseries 118} (2017) 011801},
  [\href{https://arxiv.org/abs/1606.00524}{{\ttfamily 1606.00524}}].

\bibitem{Feruglio:2017rjo}
F.~Feruglio, P.~Paradisi and A.~Pattori, \emph{{On the Importance of
  Electroweak Corrections for B Anomalies}},
  \href{https://doi.org/10.1007/JHEP09(2017)061}{\emph{JHEP} {\bfseries 09}
  (2017) 061}, [\href{https://arxiv.org/abs/1705.00929}{{\ttfamily
  1705.00929}}].

\bibitem{Crivellin:2018yvo}
A.~Crivellin, C.~Greub, D.~M\"uller and F.~Saturnino, \emph{{Importance of Loop
  Effects in Explaining the Accumulated Evidence for New Physics in B Decays
  with a Vector Leptoquark}},
  \href{https://doi.org/10.1103/PhysRevLett.122.011805}{\emph{Phys. Rev. Lett.}
  {\bfseries 122} (2019) 011805},
  [\href{https://arxiv.org/abs/1807.02068}{{\ttfamily 1807.02068}}].

\bibitem{Gherardi:2020qhc}
V.~Gherardi, D.~Marzocca and E.~Venturini, \emph{{Low-energy phenomenology of
  scalar leptoquarks at one-loop accuracy}},
  \href{https://doi.org/10.1007/JHEP01(2021)138}{\emph{JHEP} {\bfseries 01}
  (2021) 138}, [\href{https://arxiv.org/abs/2008.09548}{{\ttfamily
  2008.09548}}].

\bibitem{Crivellin:2022fdf}
A.~Crivellin, M.~Kirk, T.~Kitahara and F.~Mescia, \emph{{Large
  t\textrightarrow{}cZ as a sign of vectorlike quarks in light of the W mass}},
  \href{https://doi.org/10.1103/PhysRevD.106.L031704}{\emph{Phys. Rev. D}
  {\bfseries 106} (2022) L031704},
  [\href{https://arxiv.org/abs/2204.05962}{{\ttfamily 2204.05962}}].

\bibitem{Guedes:2022cfy}
G.~Guedes and P.~Olgoso, \emph{{A bridge to new physics: proposing new
  \textemdash{} and reviving old \textemdash{} explanations of a$_{\mu}$}},
  \href{https://doi.org/10.1007/JHEP09(2022)181}{\emph{JHEP} {\bfseries 09}
  (2022) 181}, [\href{https://arxiv.org/abs/2205.04480}{{\ttfamily
  2205.04480}}].

\bibitem{Crivellin:2023xbu}
A.~Crivellin, M.~Kirk and A.~Thapa, \emph{{Minimal model for the $W$-boson
  mass, $(g-2)_\mu$, $h\to\mu^+\mu^-$ and quark-mixing-matrix unitarity}},
  \href{https://doi.org/10.1103/PhysRevD.108.L031702}{\emph{Phys. Rev. D}
  {\bfseries 108} (2023) L031702},
  [\href{https://arxiv.org/abs/2305.03081}{{\ttfamily 2305.03081}}].

\bibitem{Lizana:2023kei}
J.~M. Lizana, J.~Matias and B.~A. Stefanek, \emph{{Explaining the $
  {B}_{d,s}\to {K}^{\left(\ast \right)}{\overline{K}}^{\left(\ast \right)} $
  non-leptonic puzzle and charged-current B-anomalies via scalar leptoquarks}},
  \href{https://doi.org/10.1007/JHEP09(2023)114}{\emph{JHEP} {\bfseries 09}
  (2023) 114}, [\href{https://arxiv.org/abs/2306.09178}{{\ttfamily
  2306.09178}}].

\bibitem{DasBakshi:2024krs}
S.~Das~Bakshi, S.~Dawson, D.~Fontes and S.~Homiller, \emph{{Relevance of
  one-loop SMEFT matching in the 2HDM}},
  \href{https://doi.org/10.1103/PhysRevD.109.075022}{\emph{Phys. Rev. D}
  {\bfseries 109} (2024) 075022},
  [\href{https://arxiv.org/abs/2401.12279}{{\ttfamily 2401.12279}}].

\bibitem{Cepedello:2024ogz}
R.~Cepedello, F.~Esser, M.~Hirsch and V.~Sanz, \emph{{Fermionic UV models for
  neutral triple gauge boson vertices}},
  \href{https://doi.org/10.1007/JHEP07(2024)275}{\emph{JHEP} {\bfseries 07}
  (2024) 275}, [\href{https://arxiv.org/abs/2402.04306}{{\ttfamily
  2402.04306}}].

\bibitem{Chala:2021pll}
M.~Chala, G.~Guedes, M.~Ramos and J.~Santiago, \emph{{Towards the
  renormalisation of the Standard Model effective field theory to dimension
  eight: Bosonic interactions I}},
  \href{https://doi.org/10.21468/SciPostPhys.11.3.065}{\emph{SciPost Phys.}
  {\bfseries 11} (2021) 065},
  [\href{https://arxiv.org/abs/2106.05291}{{\ttfamily 2106.05291}}].

\bibitem{DasBakshi:2022mwk}
S.~Das~Bakshi, M.~Chala, A.~D\'\i{}az-Carmona and G.~Guedes, \emph{{Towards the
  renormalisation of the Standard Model effective field theory to dimension
  eight: bosonic interactions II}},
  \href{https://doi.org/10.1140/epjp/s13360-022-03194-5}{\emph{Eur. Phys. J.
  Plus} {\bfseries 137} (2022) 973},
  [\href{https://arxiv.org/abs/2205.03301}{{\ttfamily 2205.03301}}].

\bibitem{DasBakshi:2023htx}
S.~Das~Bakshi and A.~D\'\i{}az-Carmona, \emph{{Renormalisation of SMEFT bosonic
  interactions up to dimension eight by LNV operators}},
  \href{https://doi.org/10.1007/JHEP06(2023)123}{\emph{JHEP} {\bfseries 06}
  (2023) 123}, [\href{https://arxiv.org/abs/2301.07151}{{\ttfamily
  2301.07151}}].

\bibitem{Chala:2023xjy}
M.~Chala and X.~Li, \emph{{Positivity restrictions on the mixing of
  dimension-eight SMEFT operators}},
  \href{https://doi.org/10.1103/PhysRevD.109.065015}{\emph{Phys. Rev. D}
  {\bfseries 109} (2024) 065015},
  [\href{https://arxiv.org/abs/2309.16611}{{\ttfamily 2309.16611}}].

\bibitem{Bakshi:2024wzz}
S.~D. Bakshi, M.~Chala, A.~D\'\i{}az-Carmona, Z.~Ren and F.~Vilches,
  \emph{{Renormalization of the SMEFT to dimension eight: Fermionic
  interactions I}}, \href{https://doi.org/10.1007/JHEP12(2024)214}{\emph{JHEP}
  {\bfseries 12} (2025) 214},
  [\href{https://arxiv.org/abs/2409.15408}{{\ttfamily 2409.15408}}].

\bibitem{Liao:2024xel}
Y.~Liao, X.-D. Ma and H.-L. Wang, \emph{{Probing dimension-8 SMEFT operators
  through neutral meson mixing}},
  \href{https://arxiv.org/abs/2409.10305}{{\ttfamily 2409.10305}}.

\bibitem{Bakshi:2021ofj}
S.~D. Bakshi, J.~Chakrabortty, C.~Englert, M.~Spannowsky and P.~Stylianou,
  \emph{{Landscaping CP-violating BSM scenarios}},
  \href{https://doi.org/10.1016/j.nuclphysb.2022.115676}{\emph{Nucl. Phys. B}
  {\bfseries 975} (2022) 115676},
  [\href{https://arxiv.org/abs/2103.15861}{{\ttfamily 2103.15861}}].

\bibitem{Ardu:2021koz}
M.~Ardu and S.~Davidson, \emph{{What is Leading Order for LFV in SMEFT?}},
  \href{https://doi.org/10.1007/JHEP08(2021)002}{\emph{JHEP} {\bfseries 08}
  (2021) 002}, [\href{https://arxiv.org/abs/2103.07212}{{\ttfamily
  2103.07212}}].

\bibitem{Allwicher:2023aql}
L.~Allwicher, G.~Isidori, J.~M. Lizana, N.~Selimovic and B.~A. Stefanek,
  \emph{{Third-family quark-lepton Unification and electroweak precision
  tests}}, \href{https://doi.org/10.1007/JHEP05(2023)179}{\emph{JHEP}
  {\bfseries 05} (2023) 179},
  [\href{https://arxiv.org/abs/2302.11584}{{\ttfamily 2302.11584}}].

\bibitem{Ciuchini:1993ks}
M.~Ciuchini, E.~Franco, G.~Martinelli, L.~Reina and L.~Silvestrini,
  \emph{{Scheme independence of the effective Hamiltonian for $b\to s\gamma$
  and $b\to s g$ decays}},
  \href{https://doi.org/10.1016/0370-2693(93)90668-8}{\emph{Phys. Lett. B}
  {\bfseries 316} (1993) 127--136},
  [\href{https://arxiv.org/abs/hep-ph/9307364}{{\ttfamily hep-ph/9307364}}].

\bibitem{Ciuchini:1993fk}
M.~Ciuchini, E.~Franco, L.~Reina and L.~Silvestrini, \emph{{Leading order QCD
  corrections to $b\to s\gamma$ and $b\to s g$ decays in three regularization
  schemes}}, \href{https://doi.org/10.1016/0550-3213(94)90223-2}{\emph{Nucl.
  Phys. B} {\bfseries 421} (1994) 41--64},
  [\href{https://arxiv.org/abs/hep-ph/9311357}{{\ttfamily hep-ph/9311357}}].

\bibitem{Buchmuller:1985jz}
W.~Buchmuller and D.~Wyler, \emph{{Effective Lagrangian Analysis of New
  Interactions and Flavor Conservation}},
  \href{https://doi.org/10.1016/0550-3213(86)90262-2}{\emph{Nucl. Phys. B}
  {\bfseries 268} (1986) 621--653}.

\bibitem{Grzadkowski:2010es}
B.~Grzadkowski, M.~Iskrzynski, M.~Misiak and J.~Rosiek, \emph{{Dimension-Six
  Terms in the Standard Model Lagrangian}},
  \href{https://doi.org/10.1007/JHEP10(2010)085}{\emph{JHEP} {\bfseries 10}
  (2010) 085}, [\href{https://arxiv.org/abs/1008.4884}{{\ttfamily 1008.4884}}].

\bibitem{Jenkins:2013zja}
E.~E. Jenkins, A.~V. Manohar and M.~Trott, \emph{{Renormalization Group
  Evolution of the Standard Model Dimension Six Operators I: Formalism and
  lambda Dependence}},
  \href{https://doi.org/10.1007/JHEP10(2013)087}{\emph{JHEP} {\bfseries 10}
  (2013) 087}, [\href{https://arxiv.org/abs/1308.2627}{{\ttfamily 1308.2627}}].

\bibitem{Jenkins:2013wua}
E.~E. Jenkins, A.~V. Manohar and M.~Trott, \emph{{Renormalization Group
  Evolution of the Standard Model Dimension Six Operators II: Yukawa
  Dependence}}, \href{https://doi.org/10.1007/JHEP01(2014)035}{\emph{JHEP}
  {\bfseries 01} (2014) 035},
  [\href{https://arxiv.org/abs/1310.4838}{{\ttfamily 1310.4838}}].

\bibitem{Alonso:2013hga}
R.~Alonso, E.~E. Jenkins, A.~V. Manohar and M.~Trott, \emph{{Renormalization
  Group Evolution of the Standard Model Dimension Six Operators III: Gauge
  Coupling Dependence and Phenomenology}},
  \href{https://doi.org/10.1007/JHEP04(2014)159}{\emph{JHEP} {\bfseries 04}
  (2014) 159}, [\href{https://arxiv.org/abs/1312.2014}{{\ttfamily 1312.2014}}].

\bibitem{Jenkins:2023bls}
E.~E. Jenkins, A.~V. Manohar, L.~Naterop and J.~Pag\`es, \emph{{Two loop
  renormalization of scalar theories using a geometric approach}},
  \href{https://doi.org/10.1007/JHEP02(2024)131}{\emph{JHEP} {\bfseries 02}
  (2024) 131}, [\href{https://arxiv.org/abs/2310.19883}{{\ttfamily
  2310.19883}}].

\bibitem{DiNoi:2024ajj}
S.~Di~Noi, R.~Gr\"ober and M.~K. Mandal, \emph{{Two-loop running effects in
  Higgs physics in Standard Model Effective Field Theory}},
  \href{https://doi.org/10.1007/JHEP12(2024)220}{\emph{JHEP} {\bfseries 12}
  (2025) 220}, [\href{https://arxiv.org/abs/2408.03252}{{\ttfamily
  2408.03252}}].

\bibitem{Henning:2014wua}
B.~Henning, X.~Lu and H.~Murayama, \emph{{How to use the Standard Model
  effective field theory}},
  \href{https://doi.org/10.1007/JHEP01(2016)023}{\emph{JHEP} {\bfseries 01}
  (2016) 023}, [\href{https://arxiv.org/abs/1412.1837}{{\ttfamily 1412.1837}}].

\bibitem{Ellis:2016enq}
S.~A.~R. Ellis, J.~Quevillon, T.~You and Z.~Zhang, \emph{{Mixed
  heavy\textendash{}light matching in the Universal One-Loop Effective
  Action}}, \href{https://doi.org/10.1016/j.physletb.2016.09.016}{\emph{Phys.
  Lett. B} {\bfseries 762} (2016) 166--176},
  [\href{https://arxiv.org/abs/1604.02445}{{\ttfamily 1604.02445}}].

\bibitem{Fuentes-Martin:2016uol}
J.~Fuentes-Martin, J.~Portoles and P.~Ruiz-Femenia, \emph{{Integrating out
  heavy particles with functional methods: a simplified framework}},
  \href{https://doi.org/10.1007/JHEP09(2016)156}{\emph{JHEP} {\bfseries 09}
  (2016) 156}, [\href{https://arxiv.org/abs/1607.02142}{{\ttfamily
  1607.02142}}].

\bibitem{Zhang:2016pja}
Z.~Zhang, \emph{{Covariant diagrams for one-loop matching}},
  \href{https://doi.org/10.1007/JHEP05(2017)152}{\emph{JHEP} {\bfseries 05}
  (2017) 152}, [\href{https://arxiv.org/abs/1610.00710}{{\ttfamily
  1610.00710}}].

\bibitem{Henning:2016lyp}
B.~Henning, X.~Lu and H.~Murayama, \emph{{One-loop Matching and Running with
  Covariant Derivative Expansion}},
  \href{https://doi.org/10.1007/JHEP01(2018)123}{\emph{JHEP} {\bfseries 01}
  (2018) 123}, [\href{https://arxiv.org/abs/1604.01019}{{\ttfamily
  1604.01019}}].

\bibitem{Kramer:2019fwz}
M.~Kr\"amer, B.~Summ and A.~Voigt, \emph{{Completing the scalar and fermionic
  Universal One-Loop Effective Action}},
  \href{https://doi.org/10.1007/JHEP01(2020)079}{\emph{JHEP} {\bfseries 01}
  (2020) 079}, [\href{https://arxiv.org/abs/1908.04798}{{\ttfamily
  1908.04798}}].

\bibitem{Cohen:2020fcu}
T.~Cohen, X.~Lu and Z.~Zhang, \emph{{Functional Prescription for EFT
  Matching}}, \href{https://doi.org/10.1007/JHEP02(2021)228}{\emph{JHEP}
  {\bfseries 02} (2021) 228},
  [\href{https://arxiv.org/abs/2011.02484}{{\ttfamily 2011.02484}}].

\bibitem{Fuentes-Martin:2020udw}
J.~Fuentes-Martin, M.~K\"onig, J.~Pag\`es, A.~E. Thomsen and F.~Wilsch,
  \emph{{SuperTracer: A Calculator of Functional Supertraces for One-Loop EFT
  Matching}}, \href{https://doi.org/10.1007/JHEP04(2021)281}{\emph{JHEP}
  {\bfseries 04} (2021) 281},
  [\href{https://arxiv.org/abs/2012.08506}{{\ttfamily 2012.08506}}].

\bibitem{Cohen:2020qvb}
T.~Cohen, X.~Lu and Z.~Zhang, \emph{{STrEAMlining EFT Matching}},
  \href{https://doi.org/10.21468/SciPostPhys.10.5.098}{\emph{SciPost Phys.}
  {\bfseries 10} (2021) 098},
  [\href{https://arxiv.org/abs/2012.07851}{{\ttfamily 2012.07851}}].

\bibitem{Dittmaier:2021fls}
S.~Dittmaier, S.~Schuhmacher and M.~Stahlhofen, \emph{{Integrating out heavy
  fields in the path integral using the background-field method: general
  formalism}},
  \href{https://doi.org/10.1140/epjc/s10052-021-09587-7}{\emph{Eur. Phys. J. C}
  {\bfseries 81} (2021) 826},
  [\href{https://arxiv.org/abs/2102.12020}{{\ttfamily 2102.12020}}].

\bibitem{Fuentes-Martin:2022jrf}
J.~Fuentes-Mart\'\i{}n, M.~K\"onig, J.~Pag\`es, A.~E. Thomsen and F.~Wilsch,
  \emph{{A proof of concept for matchete: an automated tool for matching
  effective theories}},
  \href{https://doi.org/10.1140/epjc/s10052-023-11726-1}{\emph{Eur. Phys. J. C}
  {\bfseries 83} (2023) 662},
  [\href{https://arxiv.org/abs/2212.04510}{{\ttfamily 2212.04510}}].

\bibitem{Fuentes-Martin:2023ljp}
J.~Fuentes-Mart\'\i{}n, A.~Palavri\'c and A.~E. Thomsen, \emph{{Functional
  matching and renormalization group equations at two-loop order}},
  \href{https://doi.org/10.1016/j.physletb.2024.138557}{\emph{Phys. Lett. B}
  {\bfseries 851} (2024) 138557},
  [\href{https://arxiv.org/abs/2311.13630}{{\ttfamily 2311.13630}}].

\bibitem{Fuentes-Martin:2024agf}
J.~Fuentes-Mart\'\i{}n, A.~Moreno-S\'anchez, A.~Palavri\'c and A.~E. Thomsen,
  \emph{{A Guide to Functional Methods Beyond One-Loop Order}},
  \href{https://arxiv.org/abs/2412.12270}{{\ttfamily 2412.12270}}.

\bibitem{Herzog:2017bjx}
F.~Herzog and B.~Ruijl, \emph{{The R$^{*}$-operation for Feynman graphs with
  generic numerators}},
  \href{https://doi.org/10.1007/JHEP05(2017)037}{\emph{JHEP} {\bfseries 05}
  (2017) 037}, [\href{https://arxiv.org/abs/1703.03776}{{\ttfamily
  1703.03776}}].

\bibitem{Dugan:1990df}
M.~J. Dugan and B.~Grinstein, \emph{{On the vanishing of evanescent
  operators}}, \href{https://doi.org/10.1016/0370-2693(91)90680-O}{\emph{Phys.
  Lett. B} {\bfseries 256} (1991) 239--244}.

\bibitem{Buras:1989xd}
A.~J. Buras and P.~H. Weisz, \emph{{QCD Nonleading Corrections to Weak Decays
  in Dimensional Regularization and 't Hooft-Veltman Schemes}},
  \href{https://doi.org/10.1016/0550-3213(90)90223-Z}{\emph{Nucl. Phys. B}
  {\bfseries 333} (1990) 66--99}.

\bibitem{Herrlich:1994kh}
S.~Herrlich and U.~Nierste, \emph{{Evanescent operators, scheme dependences and
  double insertions}},
  \href{https://doi.org/10.1016/0550-3213(95)00474-7}{\emph{Nucl. Phys. B}
  {\bfseries 455} (1995) 39--58},
  [\href{https://arxiv.org/abs/hep-ph/9412375}{{\ttfamily hep-ph/9412375}}].

\bibitem{Fuentes-Martin:2022vvu}
J.~Fuentes-Mart\'\i{}n, M.~K\"onig, J.~Pag\`es, A.~E. Thomsen and F.~Wilsch,
  \emph{{Evanescent operators in one-loop matching computations}},
  \href{https://doi.org/10.1007/JHEP02(2023)031}{\emph{JHEP} {\bfseries 02}
  (2023) 031}, [\href{https://arxiv.org/abs/2211.09144}{{\ttfamily
  2211.09144}}].

\bibitem{Coleman:1973jx}
S.~R. Coleman and E.~J. Weinberg, \emph{{Radiative Corrections as the Origin of
  Spontaneous Symmetry Breaking}},
  \href{https://doi.org/10.1103/PhysRevD.7.1888}{\emph{Phys. Rev. D} {\bfseries
  7} (1973) 1888--1910}.

\bibitem{Jackiw:1974cv}
R.~Jackiw, \emph{{Functional evaluation of the effective potential}},
  \href{https://doi.org/10.1103/PhysRevD.9.1686}{\emph{Phys. Rev. D} {\bfseries
  9} (1974) 1686}.

\bibitem{Iliopoulos:1974ur}
J.~Iliopoulos, C.~Itzykson and A.~Martin, \emph{{Functional Methods and
  Perturbation Theory}},
  \href{https://doi.org/10.1103/RevModPhys.47.165}{\emph{Rev. Mod. Phys.}
  {\bfseries 47} (1975) 165}.

\bibitem{Bijnens:1999hw}
J.~Bijnens, G.~Colangelo and G.~Ecker, \emph{{Renormalization of chiral
  perturbation theory to order p**6}},
  \href{https://doi.org/10.1006/aphy.1999.5982}{\emph{Annals Phys.} {\bfseries
  280} (2000) 100--139},
  [\href{https://arxiv.org/abs/hep-ph/9907333}{{\ttfamily hep-ph/9907333}}].

\bibitem{Aitchison:1984ys}
I.~J.~R. Aitchison and C.~M. Fraser, \emph{{Fermion Loop Contribution to
  Skyrmion Stability}},
  \href{https://doi.org/10.1016/0370-2693(84)90644-0}{\emph{Phys. Lett. B}
  {\bfseries 146} (1984) 63--66}.

\bibitem{Fraser:1984zb}
C.~M. Fraser, \emph{{Calculation of Higher Derivative Terms in the One Loop
  Effective Lagrangian}}, \href{https://doi.org/10.1007/BF01550255}{\emph{Z.
  Phys. C} {\bfseries 28} (1985) 101}.

\bibitem{Barvinsky:1985an}
A.~O. Barvinsky and G.~A. Vilkovisky, \emph{{The Generalized Schwinger-Dewitt
  Technique in Gauge Theories and Quantum Gravity}},
  \href{https://doi.org/10.1016/0370-1573(85)90148-6}{\emph{Phys. Rept.}
  {\bfseries 119} (1985) 1--74}.

\bibitem{Kuzenko:2003eb}
S.~M. Kuzenko and I.~N. McArthur, \emph{{On the background field method beyond
  one loop: A Manifestly covariant derivative expansion in superYang-Mills
  theories}}, \href{https://doi.org/10.1088/1126-6708/2003/05/015}{\emph{JHEP}
  {\bfseries 05} (2003) 015},
  [\href{https://arxiv.org/abs/hep-th/0302205}{{\ttfamily hep-th/0302205}}].

\bibitem{Beekveldt:2020kzk}
R.~Beekveldt, M.~Borinsky and F.~Herzog, \emph{{The Hopf algebra structure of
  the R$^{\ast}$-operation}},
  \href{https://doi.org/10.1007/JHEP07(2020)061}{\emph{JHEP} {\bfseries 07}
  (2020) 061}, [\href{https://arxiv.org/abs/2003.04301}{{\ttfamily
  2003.04301}}].

\bibitem{Chetyrkin:1997fm}
K.~G. Chetyrkin, M.~Misiak and M.~Munz, \emph{{Beta functions and anomalous
  dimensions up to three loops}},
  \href{https://doi.org/10.1016/S0550-3213(98)00122-9}{\emph{Nucl. Phys. B}
  {\bfseries 518} (1998) 473--494},
  [\href{https://arxiv.org/abs/hep-ph/9711266}{{\ttfamily hep-ph/9711266}}].

\bibitem{Collins:1994ee}
J.~C. Collins and R.~J. Scalise, \emph{{The Renormalization of composite
  operators in Yang-Mills theories using general covariant gauge}},
  \href{https://doi.org/10.1103/PhysRevD.50.4117}{\emph{Phys. Rev. D}
  {\bfseries 50} (1994) 4117--4136},
  [\href{https://arxiv.org/abs/hep-ph/9403231}{{\ttfamily hep-ph/9403231}}].

\bibitem{Naterop:2023dek}
L.~Naterop and P.~Stoffer, \emph{{Low-energy effective field theory below the
  electroweak scale: one-loop renormalization in the \textquoteright{}t
  Hooft-Veltman scheme}},
  \href{https://doi.org/10.1007/JHEP02(2024)068}{\emph{JHEP} {\bfseries 02}
  (2024) 068}, [\href{https://arxiv.org/abs/2310.13051}{{\ttfamily
  2310.13051}}].

\bibitem{Thomsen:2021ncy}
A.~E. Thomsen, \emph{{Introducing RGBeta: a Mathematica package for the
  evaluation of renormalization group $ \beta $-functions}},
  \href{https://doi.org/10.1140/epjc/s10052-021-09142-4}{\emph{Eur. Phys. J. C}
  {\bfseries 81} (2021) 408},
  [\href{https://arxiv.org/abs/2101.08265}{{\ttfamily 2101.08265}}].

\bibitem{deVries:2019nsu}
J.~de~Vries, G.~Falcioni, F.~Herzog and B.~Ruijl, \emph{{Two- and three-loop
  anomalous dimensions of Weinberg's dimension-six CP-odd gluonic operator}},
  \href{https://doi.org/10.1103/PhysRevD.102.016010}{\emph{Phys. Rev. D}
  {\bfseries 102} (2020) 016010},
  [\href{https://arxiv.org/abs/1907.04923}{{\ttfamily 1907.04923}}].

\end{thebibliography}\endgroup

\end{document}